\documentclass[journal]{IEEEtran}

\usepackage{graphicx}
\usepackage[dvipsnames]{xcolor}

\usepackage{scalerel}
\usepackage{tikz}
\usetikzlibrary{svg.path}

\definecolor{orcidlogocol}{HTML}{A6CE39}
\tikzset{
  orcidlogo/.pic={
    \fill[orcidlogocol] svg{M256,128c0,70.7-57.3,128-128,128C57.3,256,0,198.7,0,128C0,57.3,57.3,0,128,0C198.7,0,256,57.3,256,128z};
    \fill[white] svg{M86.3,186.2H70.9V79.1h15.4v48.4V186.2z}
                 svg{M108.9,79.1h41.6c39.6,0,57,28.3,57,53.6c0,27.5-21.5,53.6-56.8,53.6h-41.8V79.1z M124.3,172.4h24.5c34.9,0,42.9-26.5,42.9-39.7c0-21.5-13.7-39.7-43.7-39.7h-23.7V172.4z}
                 svg{M88.7,56.8c0,5.5-4.5,10.1-10.1,10.1c-5.6,0-10.1-4.6-10.1-10.1c0-5.6,4.5-10.1,10.1-10.1C84.2,46.7,88.7,51.3,88.7,56.8z};
  }
}

\newcommand\orcidicon[1]{\href{https://orcid.org/#1}{\mbox{\scalerel*{
\begin{tikzpicture}[yscale=-1,transform shape]
\pic{orcidlogo};
\end{tikzpicture}
}{|}}}}

\usepackage{multirow}
\usepackage{blindtext}
\usepackage{cite}
\usepackage{float}
\usepackage{stfloats}
\usepackage{hyperref}
\hypersetup{colorlinks=true,linkcolor=black,anchorcolor=black,citecolor=black,filecolor=black,menucolor=black,runcolor=black,urlcolor=black}

\graphicspath{ {figures/} }

\begin{document}

\title{Incandescent Bulb and LED Brake Lights:\\Novel Analysis of Reaction Times}

\author{ Ramaswamy Palaniappan$^{\textsuperscript{\orcidicon{0000-0001-5296-8396}}}$\ ~\IEEEmembership{Senior Member~IEEE}, Surej Mouli$^{\textsuperscript{\orcidicon{0000-0002-2876-3961}}}$\,~\IEEEmembership{Senior Member~IEEE}, Evangelina Fringi, Howard Bowman and Ian McLoughlin$^{\textsuperscript{\orcidicon{0000-0001-7111-2008}}}$\,~\IEEEmembership{Senior Member~IEEE}. 

\thanks{R. Palaniappan and E. Fringi are with the Data Science Research Group, School of Computing, University of Kent, UK (e-mail: \href{mailto:r.palani@kent.ac.uk}{r.palani@kent.ac.uk}).}
\thanks{S. Mouli is with School of Engineering \& Technology, Aston University, Birmingham, UK (e-mail: \href{mailto:r.surej@ieee.org}{surej@ieee.org}). }
\thanks{H. Bowman is with School of Computing, University of Kent and School of Psychology, University of Birmingham,UK (email:\href{mailto:H.Bowman@kent.ac.uk}{H.Bowman@kent.ac.uk}).}

\thanks{I. McLoughlin is with ICT Cluster, Singapore Institute of Technology, Singapore (email:\href{mailto:Ian.McLoughlin@singaporetech.edu.sg}{Ian.McLoughlin@singaporetech.edu.sg}).}

}

\maketitle

\begin{abstract}
Rear-end collision accounts for around $8\%$ of all vehicle crashes in the UK, with the failure to notice or react to a brake light signal being a major contributory cause. Meanwhile traditional incandescent brake light bulbs on vehicles are increasingly being replaced by a profusion of designs featuring LEDs. In this paper, we investigate the efficacy of brake light design using a novel approach to recording subject reaction times in a simulation setting using physical brake light assemblies. The reaction times of $22$ subjects were measured for ten pairs of LED and incandescent bulb brake lights. Three events were investigated for each subject, namely the latency of brake light activation to accelerator release \textit{(BrakeAcc)}, the latency of accelerator release to brake pedal depression \textit{(AccPdl)}, and the cumulative time from light activation to brake pedal depression \textit{(BrakePdl)}. To our knowledge, this is the first study in which reaction times have been split into \textit{BrakeAcc} and \textit{AccPdl}. Results indicate that the two brake lights containing incandescent bulbs led to significantly slower reaction times compared to the tested eight LED lights. \textit{BrakeAcc} results also show that experienced subjects were quicker to respond to the activation of brake lights by releasing the accelerator pedal. Interestingly, analysis also revealed that the type of brake light influenced the \textit{AccPdl} time, although experienced subjects did not always act quicker than inexperienced subjects. Overall, the study found that different designs of brake light can significantly influence driver response times.
\end{abstract}


\IEEEpeerreviewmaketitle

\section{Introduction}

\IEEEPARstart{R}{ecent} reports from the World Health Organization (WHO) have highlighted a worldwide increase in road traffic accidents, reaching $1.35$ million in $2018$ \cite{world2018global}. According to the US National Highway Traffic Safety Administration (NHTSA), rear-end crashes accounted for $7.2\%$ of total crashes in $2017$ \cite{national2017traffic}. In the same year, the Department of Transport (DoT) UK reported $13,374$ slowing or stopping related car accidents \cite{road2017}. Rear-end collisions are mostly attributed to either delayed brake response or lack of braking force due to slower reaction times, when the following drivers do not react sufficiently quickly to the behaviour of a lead vehicle due to inadequate or late detection of its deceleration \cite{winsum1996}. Many research studies have examined ways of alerting drivers to avoid rear-end crashes through improved technology either inside or outside the vehicle \cite{Bullough2007,GAO2017,ISLER2010,Li2014,LI2008}.

For example, optical looming was experimented with within a dynamic brake light system, where the brake light luminance continually and gradually expands outwards from the brake light enclosure, improving both visibility and attention of the following driver \cite{LI2008}. Stanton et al. explored a graded deceleration display technique by replacing the steady illumination of a rear centre high mounted stop lamp (CHMSL) to change brightness based on the degree of deceleration. This elicited more accurate deceleration information allowing following drivers to better gauge deceleration changes by the lead vehicle \cite{Stanton2011}. To improve the attention of the following driver, an imminent warning rear light concept was explored by Walter et al. to direct the following drivers' visual glance to the lead vehicle as it brakes rapidly to stop or slow down \cite{Wierwille2006}. Trials reported that mean brake activation time reduced from $0.35s$ for ordinary rear lighting to $0.25s$. The effectiveness of flashing brake and hazard systems in avoiding rear-end crashes was investigated by Li et al., revealing brake response time reductions of $0.14$ $\sim$  $0.62s$ for various situations tested \cite{Li2014}.

 Studies have also explored various types of stop lamps \cite{Bullough2000,bullough2001,Sivak1994}, revealing that reaction time varies by the type of lamp used in a brake light. Most automotive stop lamp types are incandescent, sweeping neon or LED. Bullough et al. evaluated these variants for CHMSLs, reporting that incandescent lamps had higher reaction times than LED or neon devices~\cite{Bullough2000}. For standard incandescent lamps, discernible optical output begins around $50$\,ms after activation, taking around $250$\,ms to reach $90\%$ of steady state output \cite{flannagan1989}. LED CHMSLs also led to shorter reaction times than neon since the high-luminance point source nature provides a stronger stimuli than the more diffused neon lamp \cite{Bullough_2001}.

 Most traffic safety studies measure driver reaction time (RT). This is a concept that traffic safety researchers have repeatedly made use of when designing experimental studies or analysing driver behaviour in crashes \cite{Summala2000,Zhang2007,MARKKULA2016}. Considering braking response, effectiveness has traditionally been measured in terms of brake reaction times (BRTs). Influential factors are usually driver age, gender, cognitive load and the various other stimuli that the driver needs to consider \cite{Barrett1968,Jeong2012,SCHWEITZER1995}. Additionally, driver reaction times differ markedly depending upon the situation; slower at lower speeds, faster in a real emergency. Their response is also affected by other issues including driver height, shoe design, pedal location, seat placement, etc. To decouple those environmental effects from the influence of the brake light design, it is necessary to separately measure how quickly a driver perceives the brake signal, and then how quickly s/he responds to it.

 As far as we are aware, there have been no extensive studies to date that used real brake lights to evaluate the effects of brake light design on the reaction time of the driver. More importantly, in this study, for the first time, we analyse reaction times by studying accelerator release timings as well as the usual brake pedal depression timings. Our experiments used ten physical brake light assemblies (two pairs containing incandescent bulbs and eight pairs containing LEDs, all of recent design) in a simulation setting, activated in a random fashion using custom built hardware.

The remainder of the paper is organised as follows. Section~\ref{sec:method} describes the experimental methodology, hardware design, data acquisition and analysis approach. Section~\ref{sec:results} presents and discusses the results and then Section~\ref{sec:conclusion} concludes the paper.

\section{Methodology}
\label{sec:method}

The experimental paradigm relied upon custom built hardware and software to present random brake light events to subjects in a simulation setting, while recording responses from a number of associated sensors.

\subsection{Experimental Setup}

The experiments were conducted in a distraction and noise-free simulation room of size $7.12 \times 14.96$m  with a projection screen at one end sized $5.00 \times 3.75$m  for replaying a highway traffic simulation video. Volunteers were seated in an automotive-style chair at a distance of $5$m  facing the screen.

\begin{figure}[tb]
\centering
\includegraphics[width=0.85\linewidth]{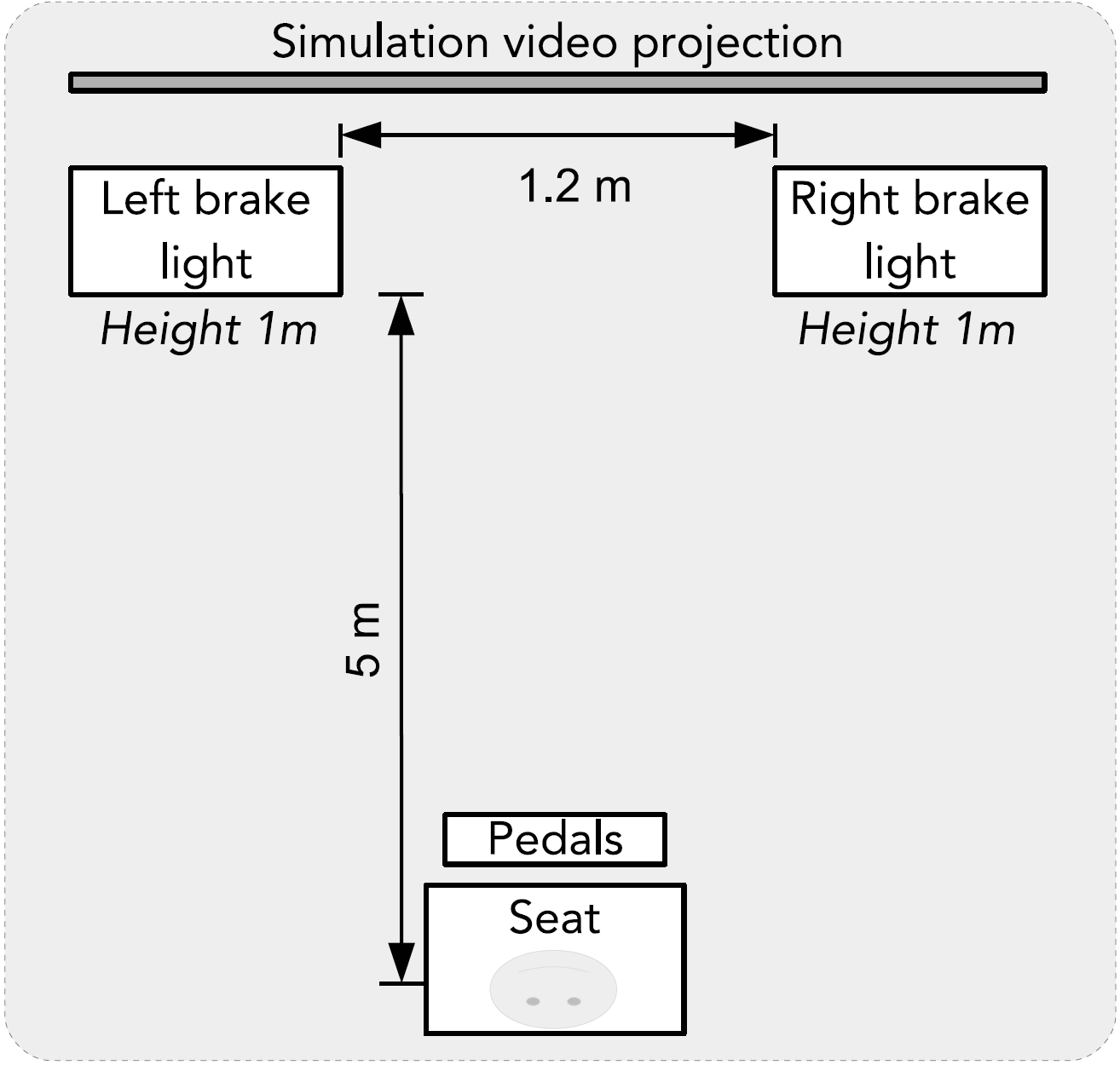}
\caption{Experimental design: (a) brake light distances (b) room layout}
\label{layout}
\end{figure}

Volunteers were provided with an accelerator and brake foot pedal assembly (QLOUNI Industrial Foot-switch Momentary Metal Foot Pedal, part number: $611702431551$), mounted in front of their seat in an arrangement as shown in Figure \ref{layout}. Custom firmware was developed to generate random braking events along with marker signals which are recorded and time stamped by an event recorder during the experiment. The firmware was programmed to generate $45$ brake light events to turn on (and then off) the brake lights, and similarly to activate the $100$mm diameter yellow distractor rings in random order. Brake light activation occurred for random periods of between $2$ and $4$s, with the distractor activation being random for between $3$ and $5$s. The control system was programmed to ensure that the distractors and brake lights were not activated simultaneously.

\subsection{Experimental Hardware}
\begin{figure}[tb]
\centering
\includegraphics[width=0.85\linewidth]{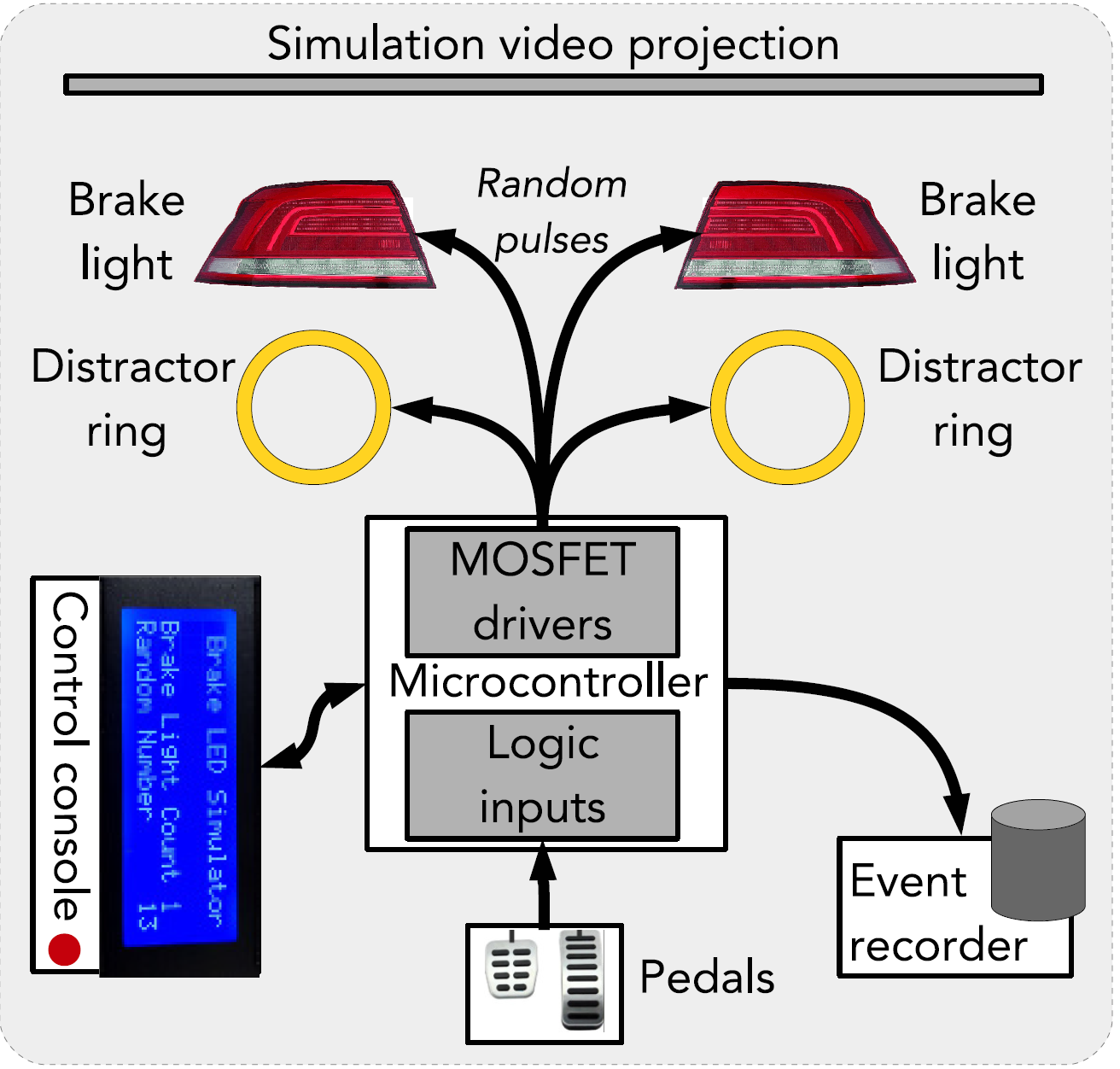}
\caption{Brake light stimulator design blocks.}
 \label{blocks}
\end{figure}

The experiment controller was designed using a custom $32$-bit microcontroller system connected to the switch sensors and a set of \textsc{MOSFET} driver circuits as shown in Figure \ref{blocks}. The control console shown was used by the person overseeing the experiments. The event recorder stored all of the timestamped information to file for later analysis.
The collected information consisted of time-stamped brake signal onset and offset times as well as onset and offset times from the two pedals.


Ten sets of brake light assemblies from different car manufacturers, selected on the basis of representing a range of distinct light shapes from common models, were used in the experiments. Table \ref{table:Table 1} lists the precise part numbers and bulb types used.

\begin{figure}[h]
\centering
\includegraphics[width=3.2in,height=3.2 in,keepaspectratio]{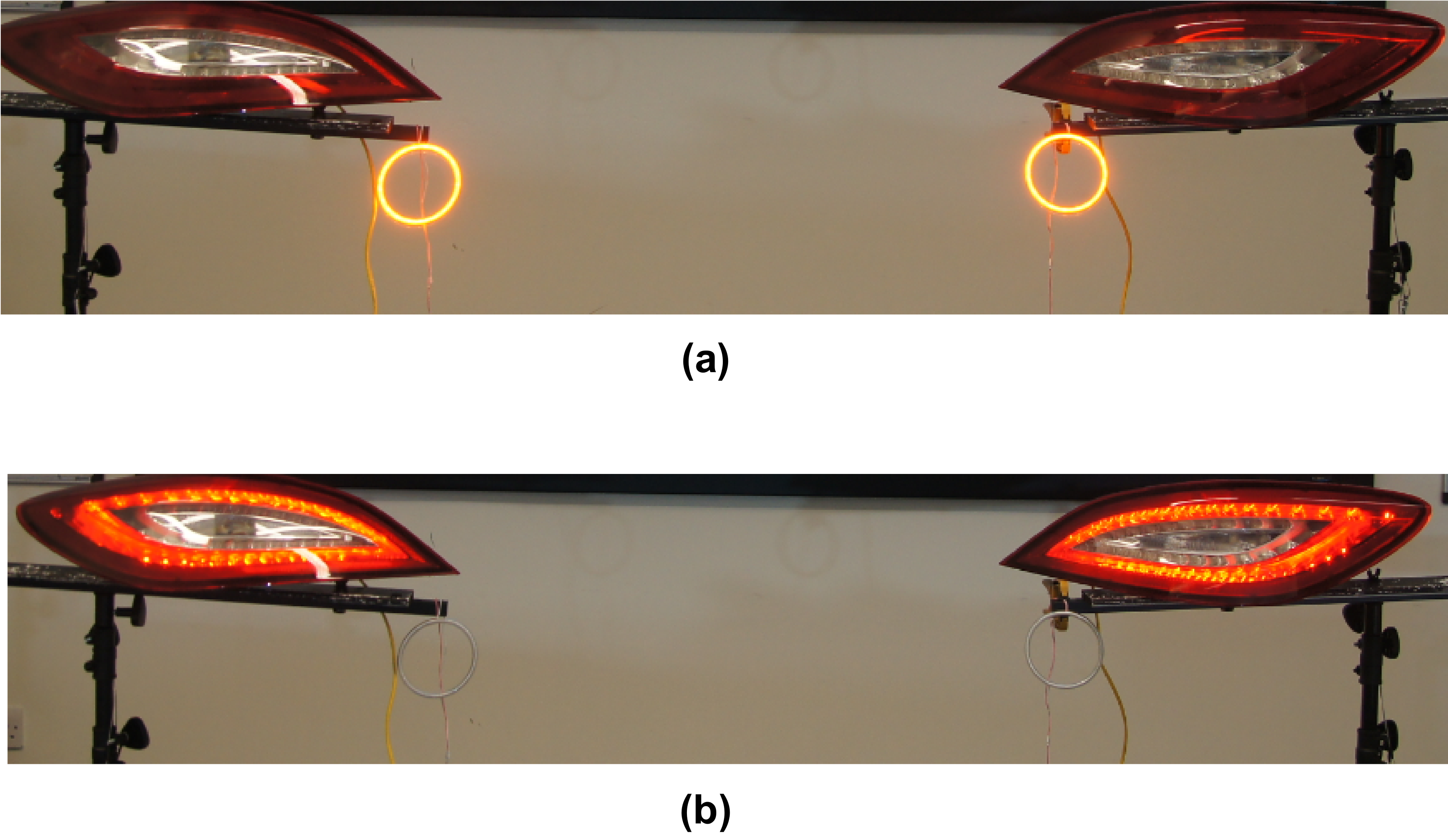}
\caption{Experimental design: (a) Yellow distractor ring with unlit Mercedes brake light (b) Mercedes brake light activated and distractor rings unlit.}
\label{mercedes}
\end{figure}

Figure \ref{mercedes} shows one of the light pairs used in this study. The brake light pairs were changed in arbitrary order between the subjects. Figure 4(a) shows the distractor rings when active while Figure 4(b) shows the activated brake light.

\begin{table}[h!]
\caption{Details of brake light assemblies used in the experiments}
\label{table:Table 1}
\centering
\begin{tabular}{lccc}
\hline
\textbf{Manufacturer} &
  \textbf{Vehicle} &
  \textbf{Part number} &
  \textbf{Bulb type}\\ \hline
Ford (Bulb) &
  \begin{tabular}[c]{@{}c@{}}Focus\\ (2018)\end{tabular} &
  \begin{tabular}[c]{@{}c@{}}1825320\\ 1825318\end{tabular} &
  Red 1490659 \\ \hline
Fiat (Bulb) &
  \begin{tabular}[c]{@{}c@{}}Fiat 500\\ (2007)\end{tabular} &
  \begin{tabular}[c]{@{}c@{}}OEN 52007424\\ OEN 52007422\end{tabular} &
  \begin{tabular}[c]{@{}c@{}}OSRAMTAIL\\ B001497\end{tabular} \\ \hline
Audi (LED) &
  \begin{tabular}[c]{@{}c@{}}Q5\\ (2016)\end{tabular} &
  \begin{tabular}[c]{@{}c@{}}8R0945093C\\ 8R0945094C\end{tabular} &
  Audi-LED \\ \hline
Fiat (LED) &
  \begin{tabular}[c]{@{}c@{}}Fiat 500\\ (2007)\end{tabular} &
  \begin{tabular}[c]{@{}c@{}}OEN 52007424\\ OEN 52007422\end{tabular} &
  82CRCANR-1 \\ \hline
Ford (LED) &
  \begin{tabular}[c]{@{}c@{}}Focus\\ (2017)\end{tabular} &
  \begin{tabular}[c]{@{}c@{}}OEN 52007424\\ OEN 52007422\end{tabular} &
  82CRCANR-1 \\ \hline
Honda (LED) &
  \begin{tabular}[c]{@{}c@{}}Civic\\ (2015)\end{tabular} &
  \begin{tabular}[c]{@{}c@{}}ULT514226\\ ULT514202\end{tabular} &
  PY21W LED \\ \hline
Mercedes (LED) &
  \begin{tabular}[c]{@{}c@{}}CLS-218\\ (2015)\end{tabular} &
  \begin{tabular}[c]{@{}c@{}}OENA2189067800\\ OEN A2189067700\end{tabular} &
  Benz-LED \\ \hline
Alfa Romeo (LED) &
  \begin{tabular}[c]{@{}c@{}}Mito\\ (2019)\end{tabular} &
  \begin{tabular}[c]{@{}c@{}}LL0604\\ LL0605\end{tabular} &
  LED P21W \\ \hline
Nissan (LED) &
  \begin{tabular}[c]{@{}c@{}}Leaf\\ (2010)\end{tabular} &
  \begin{tabular}[c]{@{}c@{}}OEN 265503NL0A\\ OEN 265553NL0A\end{tabular} &
  Nissan-LED \\ \hline
Volkswagen (LED) &
  \begin{tabular}[c]{@{}c@{}}Golf\\ (2017)\end{tabular} &
  \begin{tabular}[c]{@{}c@{}}5G0945208C\\ 5G0945207C\end{tabular} &
  VW-LED \\ \hline
\end{tabular}
\end{table}

 Eight of the brake light assemblies employed LEDs, while the remaining two sets employed incandescent bulbs. In order to make the LED/bulb comparison fairer, we included two same-vehicle model assemblies with different bulb types. Specifically, these were two sets of Ford Focus hatchback and Fiat $500$ units. The units for each vehicle had, respectively, identical exterior mouldings but employed different light technologies (i.e. there was a version using incandescent bulb and one using LED for each vehicle).

\subsection{Experimental Protocol}

The particular brake light unit pair under test were fitted to the mounts, aligned and tested. An experimental subject was then seated $5$m from the screen, as noted above in Figure~\ref{layout}. All experiments were conducted in daylight.

A motorway (UK highway) video was projected on the screen, accompanied by the natural traffic and vehicle sounds as recorded -- including tyre, engine and wind noise from the interior of the simulation vehicle as well as from passing vehicles. 
Subjects were given a task during the test with the aim of keeping their attention focused on the road.  
Specifically, they were asked to keep count of the number of times brake lights were illuminated by other vehicles during the session. 

Each session was designed as a simulated driving paradigm with the brake light assembly in front of the participant representing a leading vehicle. Those brake lights were activated at random intervals as noted above.

Subjects were instructed to continuously depress the accelerator pedal until they perceived an activation of the brake light in the simulated leading vehicle. 
At that point they were told to immediately release the accelerator and depress the brake pedal. 
They were asked to ignore any flashes or activations of the yellow distractor rings.

The experiment consisted of two sessions, taking place on separate days, each evaluating the efficacy of five different brake light configurations. The order of presentation of the lights was randomised across subjects.

Data was recorded from a total of $22$ volunteers (age $27.4\pm 5.9$ years, $M=11, F=11$). All possessed valid UK driving licenses and had normal or corrected-to-normal vision. 
Half of the subjects were classed as experienced drivers, with more than four years of driving experience. All volunteers were naive subjects recruited from the local area, and were compensated with £$100$ (£$50$ for each session) in gift vouchers for their time.

 Ethics approval for all experiments were obtained from the Faculty of Science Research Ethics committee at the University of Kent.

\subsection{Data Analysis}

 Data analysis was based on reaction time latencies evoked by the different brake lights. 
 Calculations were based on three events; the time from brake light activation to accelerator release (\textit{BrakeAcc}), the time from accelerator release to brake pedal depression (\textit{AccPdl}), and the combined brake light activation to brake pedal depression (\textit{BrakePdl}) duration.

 \textit{BrakeAcc} indicates the response time after the brake light appears and the subject releases their foot from the accelerator. This time can be considered to relate mainly to the cognitive element that starts as soon as the subject recognises the brake light, plus the time required to lift their foot from the accelerator. 
 This is followed by the more automated reflex action where the subject moves their right foot from the accelerator to depress the brake pedal. That time is denoted as \textit{AccPdl}. It is evident that the total reaction time from brake light flashing to brake pedal depression is \textit{BrakePdl} = \textit{BrakeAcc} + \textit{AccPdl}.

 As mentioned previously, each type of brake light was tested for a total of $45$ onsets for each subject, providing $180$ timing events, and thus $1800$ timing events per volunteer.

 The outputs of all analysis measures were subjected to Kruskal-Wallis tests (with \textit{$\alpha=0.05$} as significance threshold) to gauge statistical significance, since the normality of data distribution was not assumed. 
 Post-hoc Mann Whitney U testing with Bonferroni corrections were then applied where significant differences in the Kruskal-Wallis test was indicated, and thus determine any significant pair-wise differences. 
 The overall hypothesis is that more efficient brake lights will induce shorter response times (i.e. lower latencies).

\section{Results and Discussions}
\label{sec:results}

  Tables \ref{table:Table 2} and \ref{table:Table 3} present the mean $\pm$ standard deviation for \textit{BrakeAcc} and \textit{AccPdl} measurements, respectively. As can be observed from Table \ref{table:Table 2},  experienced subjects responded quicker (i.e. released the accelerator pedal faster upon seeing the brake light activation) than the inexperienced subjects. Statistically, this was different for every brake light (all pairwise cases $p<1e{-3}$) except the Fiat bulb unit ($U=-0.79, p=2.13e{-1}$). This is in line with an expectation that experienced subjects might be more subconsciously assertive to the brake signal than inexperienced subjects.

  From Table \ref{table:Table 3}, it can be seen that different brake lights also evoked different delayed responses from accelerator release to brake pedal depression (\textit{AccPdl}). The abilities of experienced vs inexperienced subjects were mixed in this regard, showing that some brake lights have an influence on the speed of the subjects’ responses while some do not. 
  Experienced subjects were quicker statistically in moving their foot from the accelerator to the brake pedal for the Ford bulb, Fiat LED and Volkswagen LED, but were slower for the Fiat bulb and Ford LED (all pairwise cases $p<3e{-1}$).

\begin{figure}[H]
\centering
\includegraphics[width=3.0in,height=3.0 in,keepaspectratio]{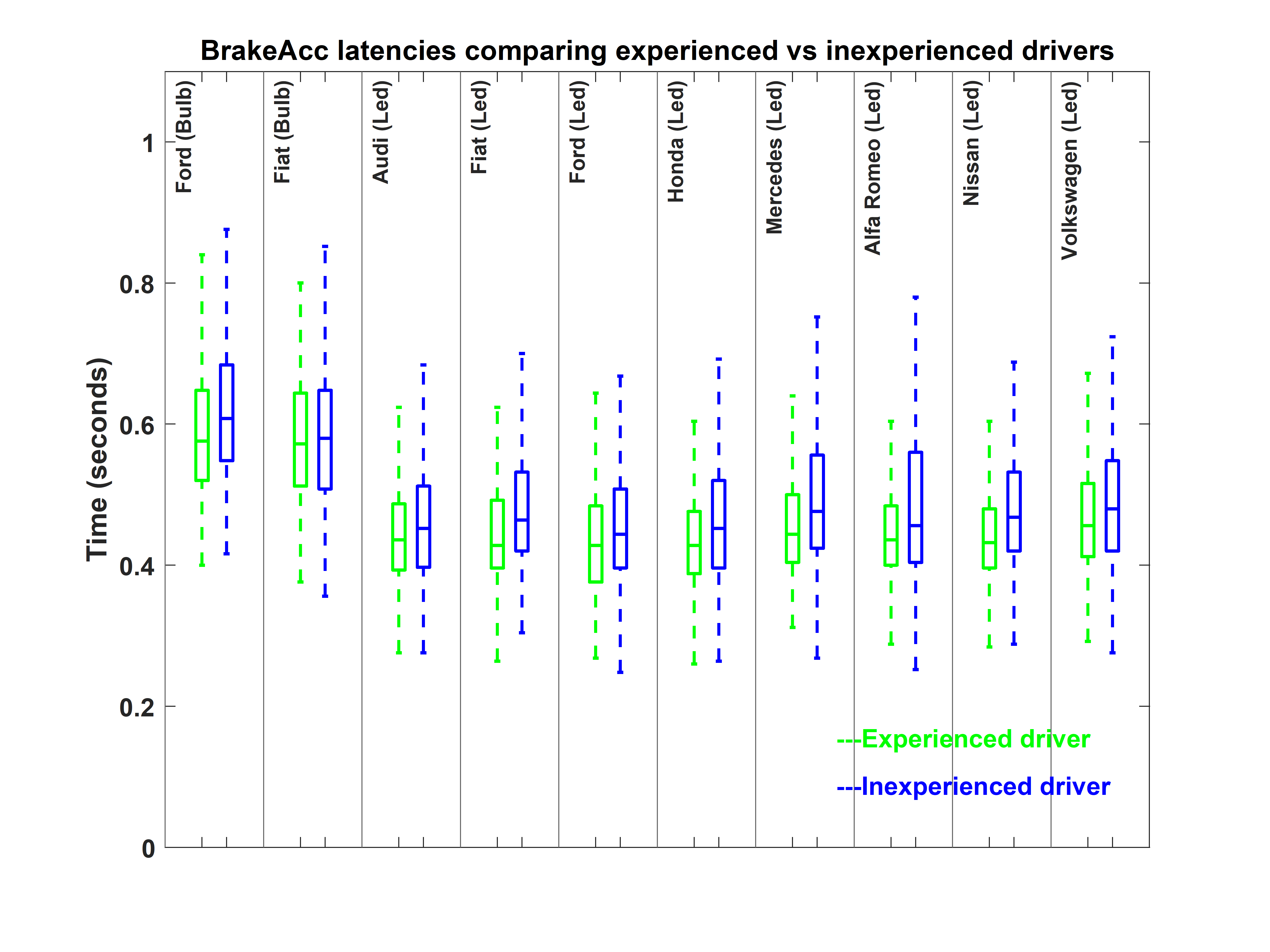}
\caption{\textit{BrakePdl} latencies comparing experienced vs inexperienced subjects.}
\label{Fig5a}
\end{figure}

\begin{figure}[tbh]
\centering
\includegraphics[width=3.0in,height=3.0 in,keepaspectratio]{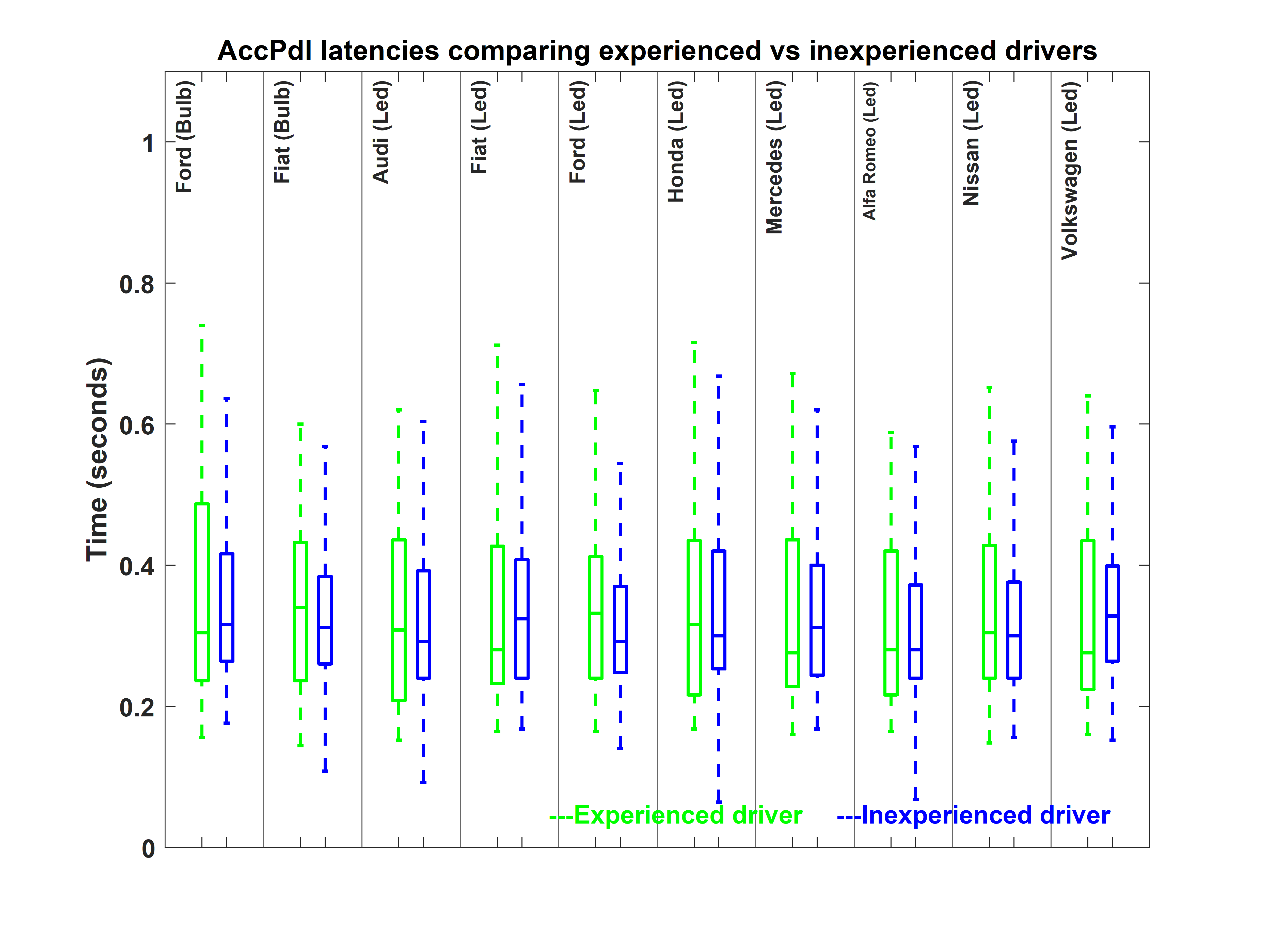}
\caption{\textit{AccPdl} latencies comparing experienced vs inexperienced subjects.}
\label{Fig6a}
\end{figure}

\begin{figure}[tbh]
\centering
\includegraphics[width=3.0in,height=3.0 in,keepaspectratio]{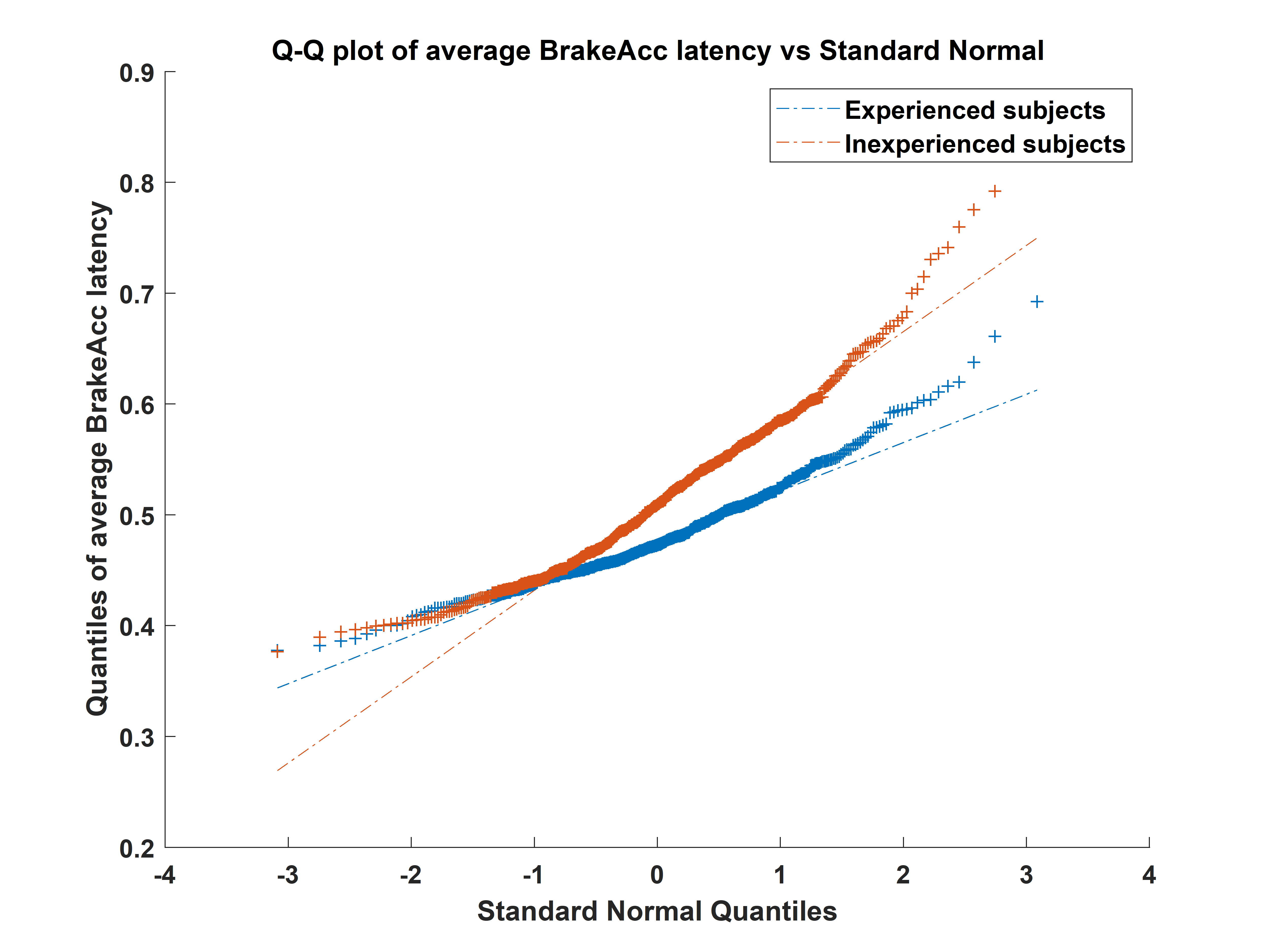}
\caption{Accelerator release latency \textit{(BrakePdl)} subject wise for ALL brake lights.}
\label{Fig7a}
\end{figure}

\begin{figure}[tbh]
\centering
\includegraphics[width=3.0in,height=3.0 in,keepaspectratio]{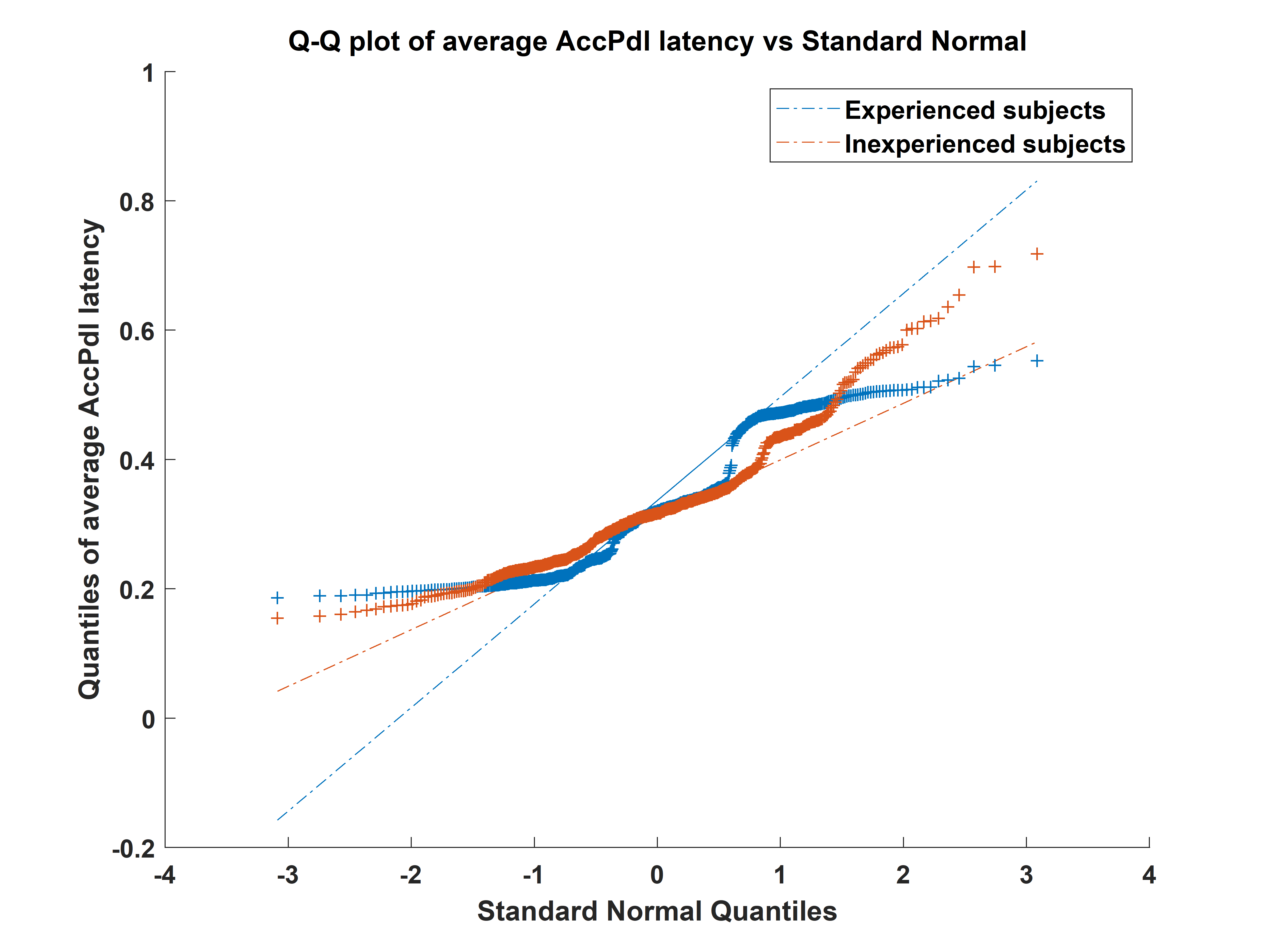}
\caption{Accelerator release to brake pedal latency \textit{(AccPdl)} subject wise for ALL brake lights.}
\label{Fig8a}
\end{figure}

  Figures \ref{Fig5a} and \ref{Fig6a} show boxplots of latencies for \textit{BrakeAcc} and \textit{AccPdl}, respectively. 
  It is evident from the figures that the median values of \textit{BrakeAcc} were smaller for experienced subjects compared to inexperienced ones, which was true for every brake light (for \textit{AccPdl}, it was mixed though). 
  Figures \ref{Fig7a} and \ref{Fig8a} show the quantile-quantile (Q-Q) plot for \textit{BrakeAcc} and \textit{AccPdl} latencies for experienced vs inexperienced subjects. 
  It can be seen that the inexperienced subjects had much longer \textit{BrakeAcc} distributions (the distributions are similar early on, but diverge later). 
  This showed that their overall medians were longer than for experienced subjects, but more importantly at the slow end of the reaction time distribution, inexperienced subjects were especially slow. 
  This \textit{slowness} in response is very important as it could be a causal factor in accidents; where drivers are slow to respond and thus crash into the the car in front. 
  However, this difference was not clearly evident for \textit{AccPdl} latency, despite divergence later on showing the slowness of response for inexperienced subjects.

\begin{figure}[tbh]
\centering
\includegraphics[width=3.0in,height=3.0 in,keepaspectratio]{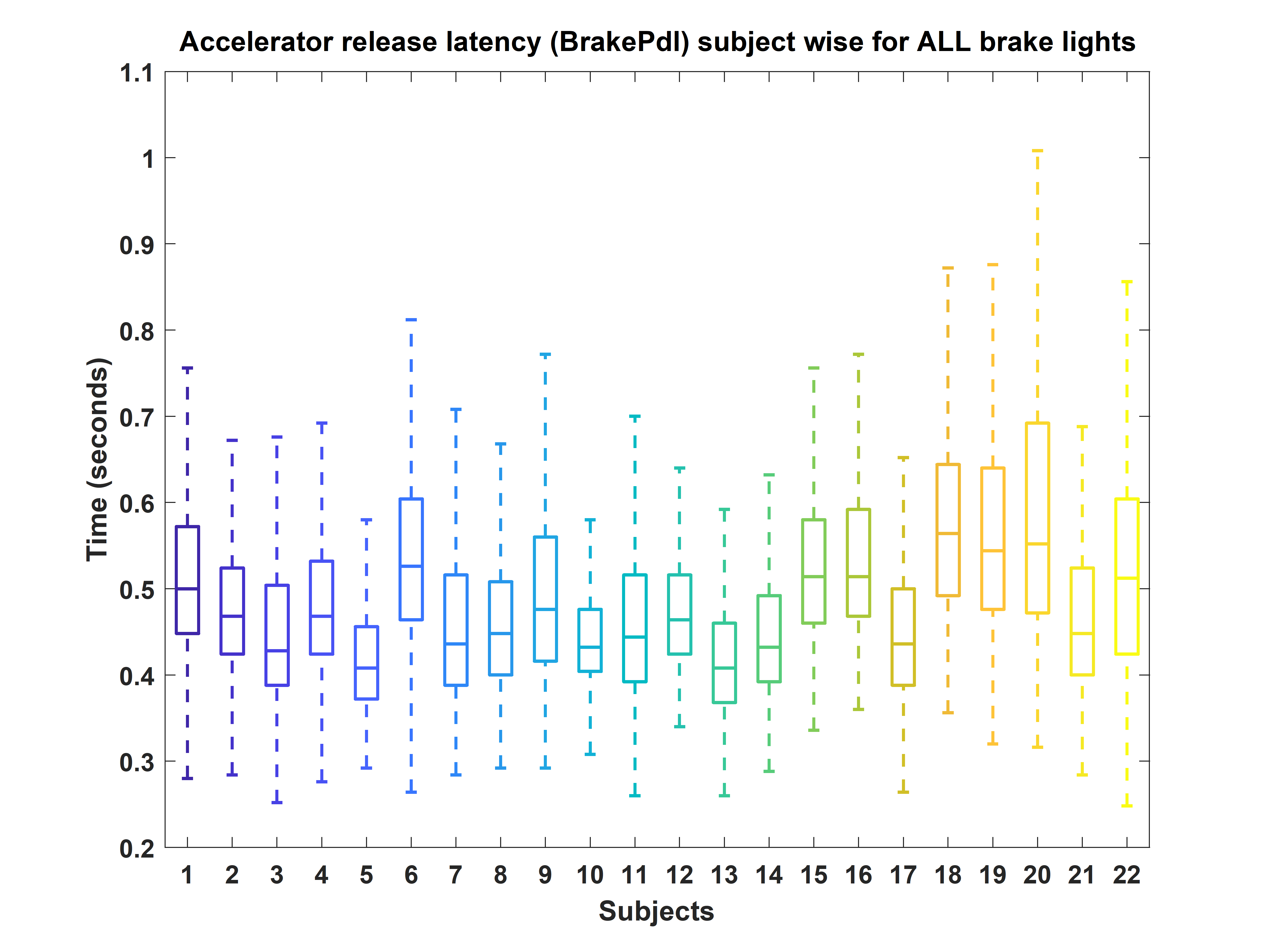}
\caption{Accelerator release latency \textit{(BrakePdl)} subject wise for ALL brake lights (the first 11 subjects were experienced drivers).}
\label{Fig9a}
\end{figure}

\begin{figure}[tbh]
\centering
\includegraphics[width=3.0in,height=3.0 in,keepaspectratio]{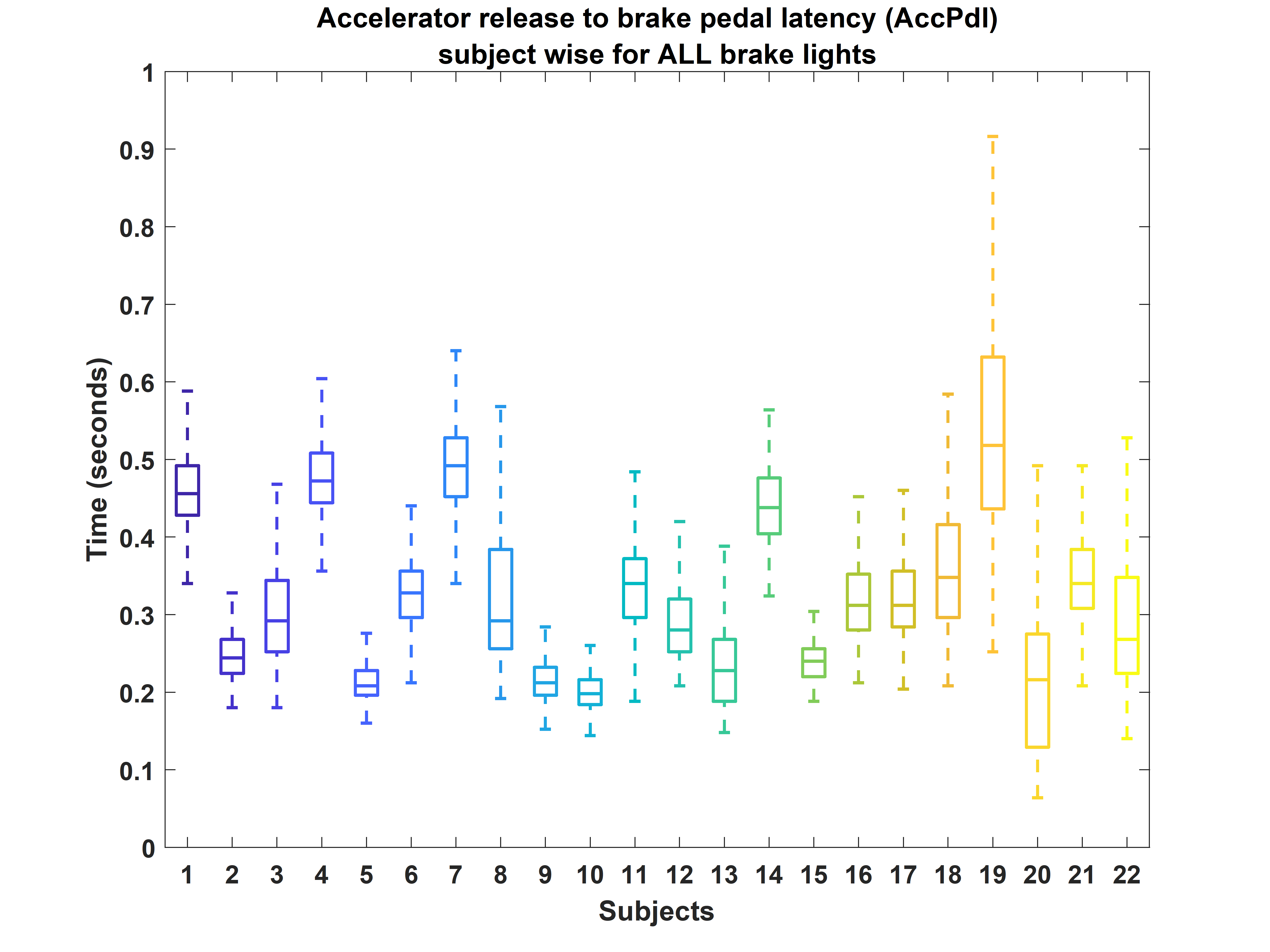}
\caption{Accelerator release to brake pedal latency \textit{(AccPdl)} subject wise for ALL brake lights (the first 11 subjects were experienced drivers).}
\label{Fig10a}
\end{figure}

 Comparing all the subjects (as shown in Figures \ref{Fig9a} and \ref{Fig10a}), a statistical difference was also noted for both response latencies showing that subjects' responses were dissimilar: \textit{BrakeAcc}: $(H(9)=2352.05, p=0)$, \textit{AccPdl}: $(H(9)=46.91, p=4.08e{-7})$. The first $11$ shown in the figures were experienced subjects with the rest being inexperienced. 

\begin{figure}[tbh]
\centering
\includegraphics[width=3.0in,height=3.0 in,keepaspectratio]{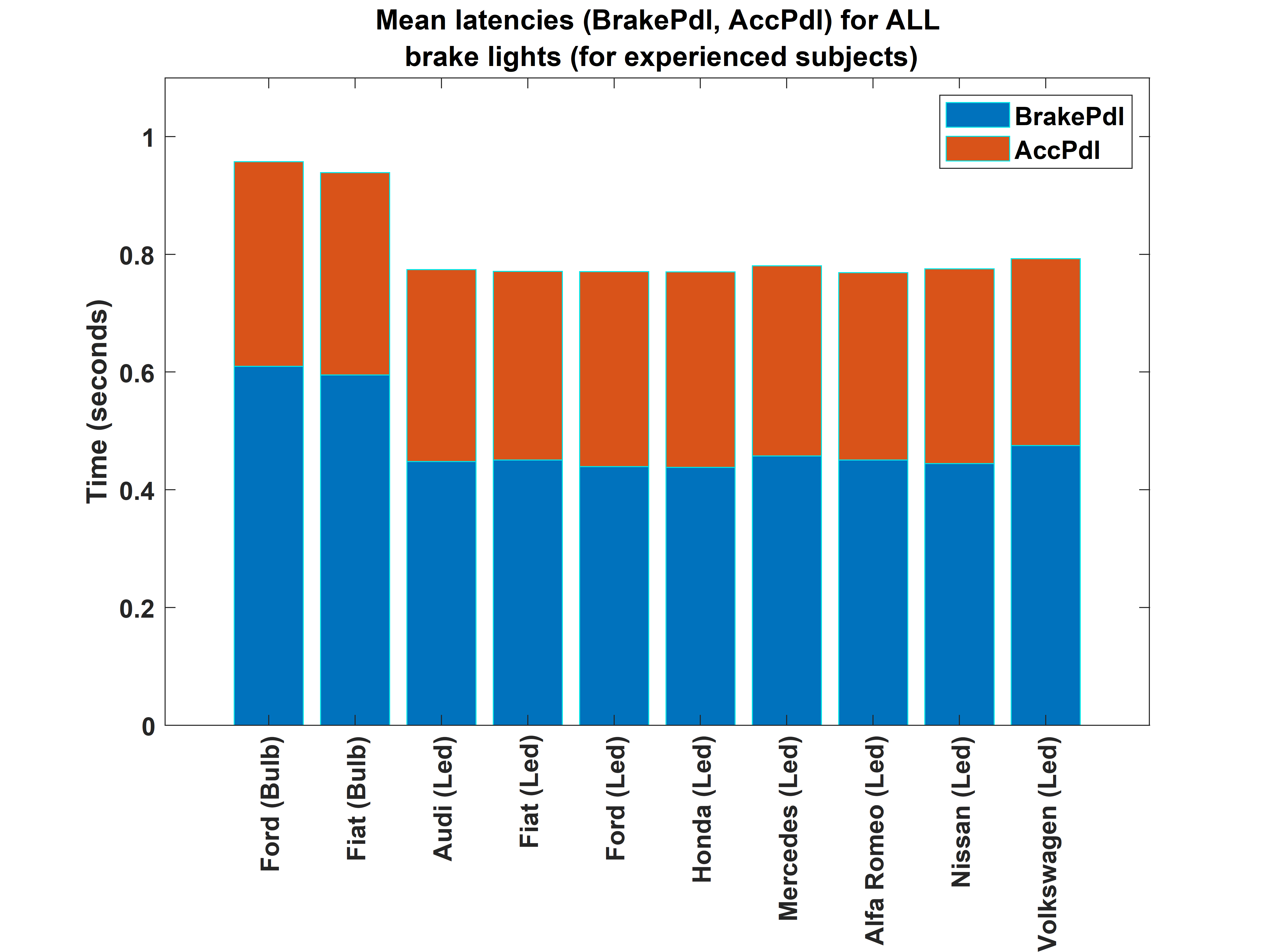}
\caption{Mean latencies \textit{(BrakePdl, AccPdl)} for ALL brake lights (for the experienced drivers).}
\label{Fig11a}
\end{figure}

  Brake light reaction times for the $11$ experienced subjects based on \textit{BrakeAcc} and \textit{AccPdl} are shown in Figure \ref{Fig11a}. As can be seen from the plot (the blue portion of the bars), both bulb versions of the brake assemblies from Ford and Fiat have the highest \textit{BrakeAcc} response times (which was statistically significant from the eight LED lights, $(H(8), p=0)$ denoting that they were the slowest lights to draw a response. Between the two bulb units, there was no significant difference statistically ($U=-1.38, p=0.16$). Among the LED brake lights, the slowest (i.e. the highest latency) was from Volkswagen which was statistically significant from every other LED light (all pairwise cases $p<1e{-9}$), while the next slowest was Mercedes -- however this was significant only compared to the Ford ($U=-3.84, p=6.11e{-5}$) and Honda ($U=-3.81, p=7.05e{-5}$) units. 
  The lowest \textit{BrakeAcc} latency (i.e. the fastest light) were the Ford, Honda and Nissan units, although only Volkswagen and Mercedes indicated statistically significant differences in terms of the slower latencies as mentioned. 
  This could possibly be due to their distinct characteristics: the Ford LED having the largest lit area, the Honda LED being the brightest, and the Nissan unit having the longest vertical lit dimension. 
  Our previous studies based on brain response to LED light shapes revealed significant influence on cognitive responses for various shapes, orientations and brightness \cite{mouli2013performance,Mouli2016Elic, Mouli2015Quantification}.

 Considering the reaction times for the $11$ experienced subjects based on \textit{AccPdl} responses (the red portion of the bar), the general thought is that there should be no difference in terms of \textit{AccPdl}. It should be relatively constant for each subject. However, the results indicated otherwise. 
 The Ford bulb timings were significantly slower than for the Audi  ($U=-2.86, p=2.10e{-3}$), Alfa Romeo  ($U=-3.72, p=1.01e{-4}$) and Volkswagen ($U=-3.63, p=1.41e{-4}$) LED units. 
 Meanwhile the Fiat bulb timings were slower than the Audi ($U=-2.84, p=2.20e{-3}$) and Alfa Romeo LED lights ($U=-3.74, p=9.21e{-5}$). 
 This indicated that the bulb had an additional negative effect which acted to reduce the reflex response component, in addition to the cognitive component. 
 While we are analysing this effect further, we conjecture that the shape and/or illumination level influences not only how quickly a subject can detect the brake signal, but how tentative or decisive the consequent response is.

 Considering the total reaction time, \textit{BrakePdl} (as shown in Figure $11$, the full bars, both blue and red sections), in line with the other results, reports both the bulbs being statistically slower than any of the LED lights $(H(8), p=0)$. However, within the LED lights, there was no statistically significant difference between units $(H(7)=4.99, p=0.66)$. However, from the plot we can see that Volkswagen LED tended to be the slowest, followed by the Mercedes unit.

\begin{figure}[tbh]
\centering
\includegraphics[width=3.0in,height=3.0 in,keepaspectratio]{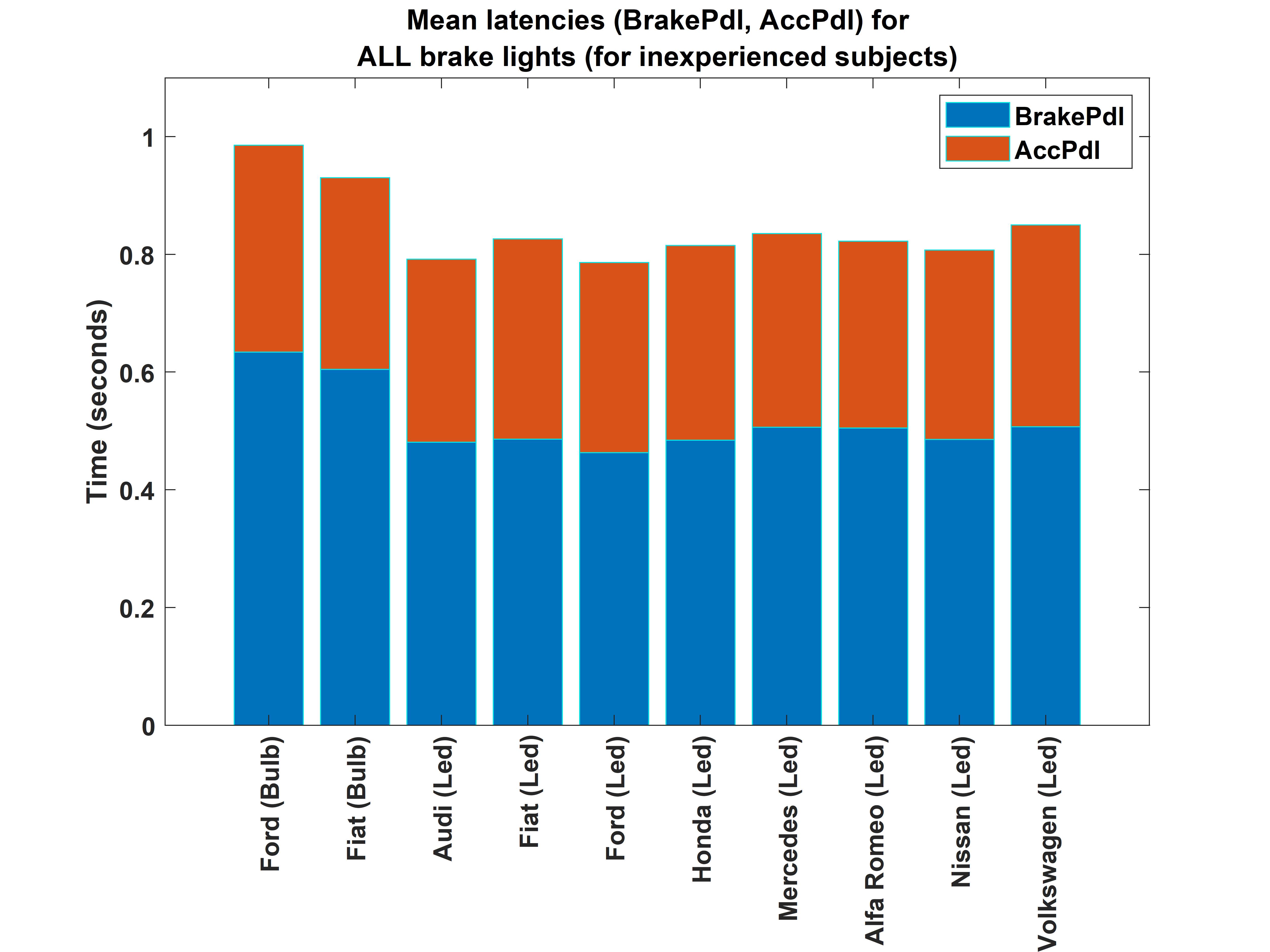}
\caption{Mean latencies \textit{(BrakePdl, AccPdl)} for ALL brake lights (for the inexperienced drivers).}
\label{Fig12a}
\end{figure}

The\textit{ BrakeAcc} responses from the inexperienced subjects is shown in Figure \ref{Fig12a}  (as the blue portion of the bars). 
 The slowest responses were from both the bulbs  ($H(8), p=0$); between the bulb assemblies, the Ford was slower than the Fiat, ($U=4.49, p=3.62e{-6}$). 
 Among the LED units, the slowest was from Volkswagen (statistically significant against all other LED units,$(H(7)=72.83, p=3.96e{-13})$. This was followed by the Mercedes, which was statistically slower than the Audi ($U=4.94, p=3.97e{-7}$), Ford LED ($U=6.38, p=9.14e{-11}$) and Honda LED ($U=4.78, p=8.61e{-7}$). 
 The fastest light was the Ford LED (which was statistically different from all but the Audi ($U=1.19, p=1.17e{-1}$) and the Honda unit ($U=1.36, p=8.66e{-2}$)).

In terms of \textit{AccPdl} (shown in Figure \ref{Fig12a} as the red portion of the bars), the expectation is again that there should not be any difference between the lights since the reflex response is what is being analysed. 
However both the Ford bulb and Volkswagen LED are statistically slower than the Alfa Romeo and Nissan LED units (all pairwise cases $p<1e{-5}$). 

As expected from analysis of \textit{BrakeAcc} and \textit{AccPdl}, both the bulbs were slower than any of the LED lights when considering the total reaction times (the full bars in Figure \ref{Fig12a}) ($H(8), p=0$). Among the LED units, there were more differences exhibited than there were for the experienced subjects. 
For example, the Volkswagen was statistically slower than the Audi, Nissan, Alfa Romeo, Ford, and Honda units (all pairwise $p<1e{-4}$). 
Meanwhile the Ford LED unit was quicker statistically than those from Mercedes and Fiat (all pairwise $p<1e{-4}$).

\begin{figure}[h!]
\centering
\includegraphics[width=3.0in,height=3.0 in,keepaspectratio]{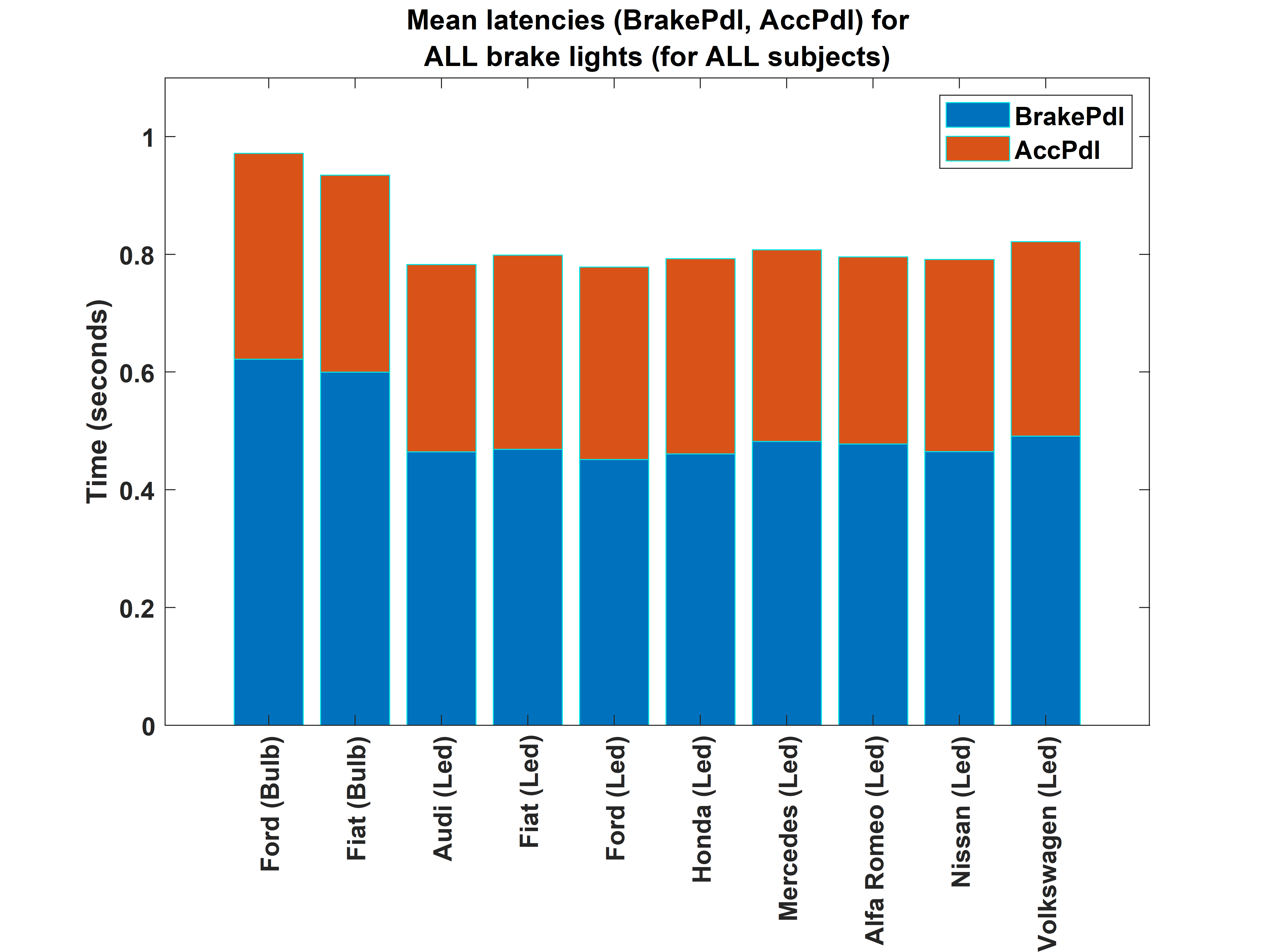}
\caption{Mean latencies \textit{(BrakePdl, AccPdl)} for ALL brake lights (for ALL subjects).}
\label{Fig13a}
\end{figure}

Figure \ref{Fig13a} compares the \textit{BrakeAcc, AccPdl} and \textit{BrakePdl} results (blue, red, full bars, respectively) for each brake light for all $22$ subjects combined. 
Combining the \textit{BrakePdl} analyses from both experienced and inexperienced subjects, both the bulbs are slower statistically than any of the LED lights $(H(8), p=0)$. 
The fastest LED was from Ford -- statistically significant against all LED lights except the Audi and Honda  (all pairwise $p<1e{-3}$). 
The slowest was the Volkswagen unit (statistically significant from all other LED lights($H(7)=124.23, p=1.00e{-23}$), followed by the Mercedes LED (though it is statistically significant from the Audi, Ford, Honda and Nissan LED lights only (all pairwise $p<1e{-4}$)).

\begin{table*}[tb]
\centering
\caption{Mean latency and standard deviation (\textit{BrakeAcc}, \textit{AccPdl}, in seconds) for each brake light from all subjects}
\label{table:Table 2}
\resizebox{\textwidth}{!}{%
\begin{tabular}{l | c c | c c c c c c c c}
 &
  \multicolumn{2}{c|}{\textbf{Bulb}} &
  \multicolumn{8}{c}{\textbf{LED}} \\

 & Ford & Fiat & Audi & Fiat & Ford & Honda & Mercedes & Alfa Romeo & Nissan & Volkswagen \\ \hline
\textbf{Subject} & \multicolumn{10}{l}{\textit{BrakeAcc}} \\ \hline
Experienced & 
  $0.61\pm0.17$ &
  $0.59\pm0.15$ &
  $0.45\pm0.08$ &
  $0.45\pm0.10$ &
  $0.44\pm0.09$ &
  $0.44\pm0.08$ &
  $0.46\pm0.08$ &
  $0.45\pm0.08$ &
  $0.44\pm0.08$ &
  $0.47\pm0.11$ \\
Inexperienced &
  $0.63\pm0.16$ &
  $0.60\pm0.17$ &
  $0.48\pm0.15$ &
  $0.49\pm0.11$ &
  $0.46\pm0.12$ &
  $0.48\pm0.17$ &
  $0.51\pm0.14$ &
  $0.50\pm0.18$ &
  $0.49\pm0.11$ &
  $0.51\pm0.15$ \\
All subjects &
  $0.62\pm0.16$ &
  $0.60\pm0.16$ &
  $0.46\pm0.12$ &
  $0.47\pm0.10$ &
  $0.45\pm0.11$ &
  $0.46\pm0.13$ &
  $0.48\pm0.12$ &
  $0.48\pm0.14$ &
  $0.46\pm0.10$ &
  $0.49\pm0.13$ \\ \hline
\textbf{Subject} & \multicolumn{10}{l}{\textit{AccPdl}} \\ \hline
Experienced &
  $0.35\pm0.14$ &
  $0.34\pm0.11$ &
  $0.33\pm0.12$ &
  $0.32\pm0.11$ &
  $0.33\pm0.10$ &
  $0.33\pm0.12$ &
  $0.32\pm0.12$ &
  $0.32\pm0.11$ &
  $0.33\pm0.11$ &
  $0.32\pm0.12$ \\
Inexperienced &
  $0.35\pm0.13$ &
  $0.33\pm0.17$ &
  $0.31\pm0.11$ &
  $0.34\pm0.12$ &
  $0.32\pm0.13$ &
  $0.33\pm0.15$ &
  $0.33\pm0.11$ &
  $0.32\pm0.16$ &
  $0.32\pm0.13$ &
  $0.34\pm0.12$ \\
All subjects &
  $0.35\pm0.13$ &
  $0.33\pm0.15$ &
  $0.32\pm0.11$ &
  $0.33\pm0.12$ &
  $0.33\pm0.11$ &
  $0.33\pm0.14$ &
  $0.33\pm0.11$ &
  $0.32\pm0.14$ &
 $0.33\pm0.12$ &
 $0.33\pm0.12$ \\
\end{tabular}%
}
\end{table*}

\begin{table}[tb]
\centering
\caption{Mean latency and standard deviation (\textit{BrakeAcc, AccPdl}, in seconds) from all brake lights for experienced drivers 1--11 (top) and inexperienced drivers 12--22 (bottom).}
\label{table:Table 3}
\begin{tabular}{lcc}
\hline
 \textbf{Subject}  & \textit{BrakeAcc}  & \textit{AccPdl}     \\ \hline
1       & $0.52\pm0.10$ & $0.46\pm0.05$ \\
2       & $0.49\pm0.11$ & $0.25\pm0.04$ \\
3       & $0.46\pm0.13$ & $0.31\pm0.07$ \\
4       & $0.49\pm0.10$ & $0.47\pm0.06$ \\
5       & $0.42\pm0.08$ & $0.22\pm0.05$ \\
6       & $0.55\pm0.13$ & $0.34\pm0.07$ \\
7       & $0.47\pm0.12$ & $0.49\pm0.07$ \\
8       & $0.47\pm0.14$ & $0.32\pm0.09$ \\
9       & $0.51\pm0.15$ & $0.22\pm0.05$ \\
10      & $0.45\pm0.07$ & $0.20\pm0.03$ \\
11      & $0.47\pm0.14$ & $0.34\pm0.06$ \\
Average exp.    & $0.48\pm0.03$ & $0.33\pm 0.11$ \\ \hline
12      & $0.48\pm0.10$ & $0.29\pm0.06$ \\
13      & $0.42\pm0.09$ & $0.24\pm0.06$ \\
14      & $0.45\pm0.12$ & $0.44\pm0.06$ \\
15      & $0.53\pm0.11$ & $0.24\pm0.04$ \\
16      & $0.55\pm0.17$ & $0.32\pm0.05$ \\
17      & $0.46\pm0.11$ & $0.32\pm0.06$ \\
18      & $0.57\pm0.11$ & $0.37\pm0.09$ \\
19      & $0.58\pm0.21$ & $0.55\pm0.17$ \\
20      & $0.60\pm0.18$ & $0.22\pm0.11$ \\
21      & $0.48\pm0.16$ & $0.35\pm0.07$ \\
22      & $0.54\pm0.19$ & $0.30\pm0.14$ \\
Average inexp.   & $0.52\pm0.06$ & $0.33\pm0.10$  \\ \hline 
Overall average & $0.50\pm0.05$ &$0.33\pm0.10$ \\ 
\end{tabular}
\end{table}

\begin{table*}[tb]
\centering
\caption{LED brake lights with the fastest and slowest response times for all subjects}
\label{table:Table 4}
\begin{tabular}{c|llllll}
 &
  \multicolumn{2}{c}{\textbf{\textit{BrakeAcc }latency}} &
  \multicolumn{2}{c}{\textbf{\textit{AccPdl} latency}} &
  \multicolumn{2}{c}{\textbf{\textit{BrakePdl} latency}} \\ 
Subject & Fastest LED & Slowest LED & Fastest LED & Slowest LED & Fastest LED & Slowest LED\\
 \hline
 1 & 
 { Ford} &
  { Volkswagen} &
  { Alfa Romeo} &
  { Volkswagen} &
  { Ford} &
  { Volkswagen} \\
2 &
  { Fiat} &
  { Ford} &
  { Alfa Romeo} &
  { Ford} &
  { Fiat} &
  { Ford} \\
3 &
  { Ford} &
  { Mercedes} &
  { Volkswagen} &
  { Honda} &
  { Volkswagen} &
  { Honda} \\
4 &
  { Mercedes} &
  { Honda} &
  { Ford} &
  { Fiat} &
  { Ford} &
  { Nissan} \\ 
5 &
  { Honda} &
  { Volkswagen} &
  { Audi} &
  { Nissan} &
  { Honda} &
  { Nissan} \\ 
6 &
  { Nissan} &
  { Alfa Romeo} &
  { Alfa Romeo} &
  { Alfa Romeo} &
  { Honda} &
  { Alfa Romeo} \\ 
7 &
  { Ford} &
  { Volkswagen} &
  { Ford} &
  { Honda} &
  { Ford} &
  { Volkswagen} \\ 
8 &
  { Honda} &
  { Fiat} &
  { Mercedes} &
  { Honda} &
  { Mercedes} &
  { Ford} \\ 
9 &
  { Ford} &
  { Fiat} &
  { Audi} &
  { Mercedes} &
  { Audi} &
  { Fiat} \\ 
10 &
  { Alfa Romeo} &
  { Volkswagen} &
  { Volkswagen} &
  { Ford} &
  { Audi} &
  { Ford} \\ 
11 &
  { Honda} &
  { Ford} &
  { Mercedes} &
  { Ford} &
  { Mercedes} &
  { Ford} \\
12 &
  { Fiat} &
  { Mercedes} &
  { Alfa Romeo} &
  { Audi} &
  { Fiat} &
  { Audi} \\ 
13 &
  { Alfa Romeo} &
  { Mercedes} &
  { Audi} &
  { Mercedes} &
  { Audi} &
  { Mercedes} \\ 
14 &
  { Ford} &
  { Mercedes} &
  { Nissan} &
  { Honda} &
  { Ford} &
  { Fiat} \\ 
15 &
  { Fiat} &
  { Mercedes} &
  { Fiat} &
  { Alfa Romeo} &
  { Fiat} &
  { Mercedes} \\ 
16 &
  { Fiat} &
  { Honda} &
  { Honda} &
  { Fiat} &
  { Audi} &
  { Volkswagen} \\ 
17 &
  { Audi} &
  { Mercedes} &
  { Audi} &
  { Volkswagen} &
  { Audi} &
  { Mercedes} \\
18 &
  { Ford} &
  { Mercedes} &
  { Ford} &
  { Honda} &
  { Ford} &
  { Nissan} \\
19 &
  { Honda} &
  { Alfa Romeo} &
  { Audi} &
  { Alfa Romeo} &
  { Audi} &
  { Alfa Romeo} \\ 
20 &
  { Volkswagen} &
  { Alfa Romeo} &
  { Alfa Romeo} &
  { Volkswagen} &
  { Ford} &
  { Mercedes} \\ 
21 &
  { Honda} &
  { Volkswagen} &
  { Ford} &
  { Mercedes} &
  { Ford} &
  { Volkswagen} \\ 
22 &
  { Alfa Romeo} &
  { Volkswagen} &
  { Ford} &
  { Volkswagen} &
  { Ford} &
  { Volkswagen} \\ 
\end{tabular}

\end{table*}

 Even though this study was focused on the cognitive response invoked by the various brake lights, interestingly the brake lights also influenced the reflex time taken for the foot to release from the accelerator and depress the brake pedal. 
 Combining \textit{AccPdl} from both experienced and inexperienced subjects, the Ford bulb was statistically slower than the Alfa Romeo, Mercedes, Audi and Nissan LED lights (all pairwise $p<1e{-4}$) while the Fiat bulb was slower than the Audi and Alfa Romeo LED lights (all pairwise $p<1e{-3}$). 
 Among the LED lights, there were some significant differences ($H(7)=16.5, p=2.09e{-1}$) with the Alfa Romeo being faster than the Honda ($U=2.86, p=2e{-3}$) and the Volkswagen units ($U=-3.26, p=5.64e{-4}$).

\begin{figure}[tbh]
\centering
\includegraphics[width=3.0in,height=3.0 in,keepaspectratio]{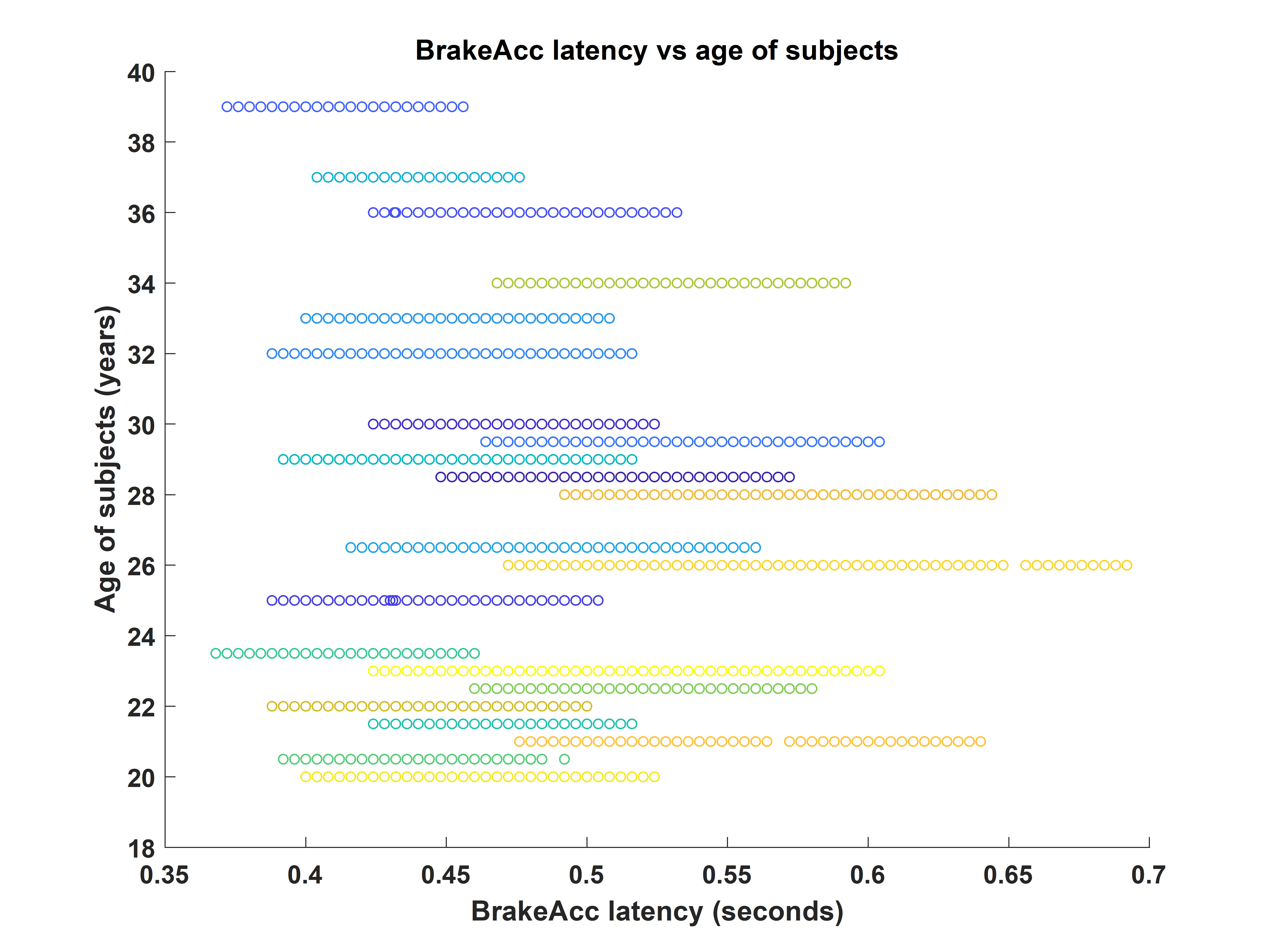}
\caption{\textit{BrakeAcc }latency with respect to subject age.}
\label{Fig14a}
\end{figure}
\begin{figure}[tbh]
\centering
\includegraphics[width=3.0in,height=3.0 in,keepaspectratio]{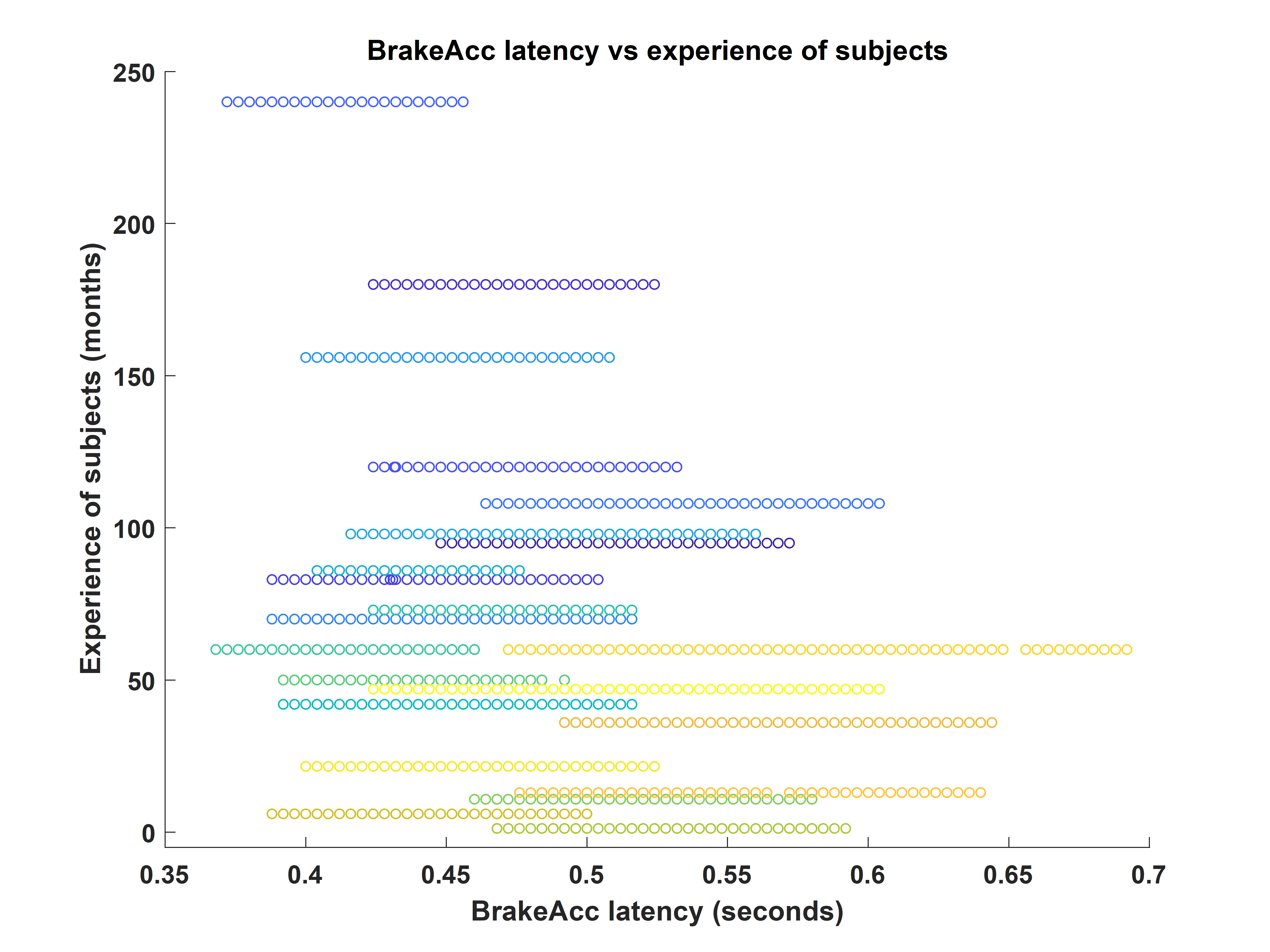}
\caption{\textit{BrakeAcc }latency vs experience level of subjects.}
\label{Fig15a}
\end{figure}
\begin{figure}[tbh]
\centering
\includegraphics[width=3.0in,height=3.0 in,keepaspectratio]{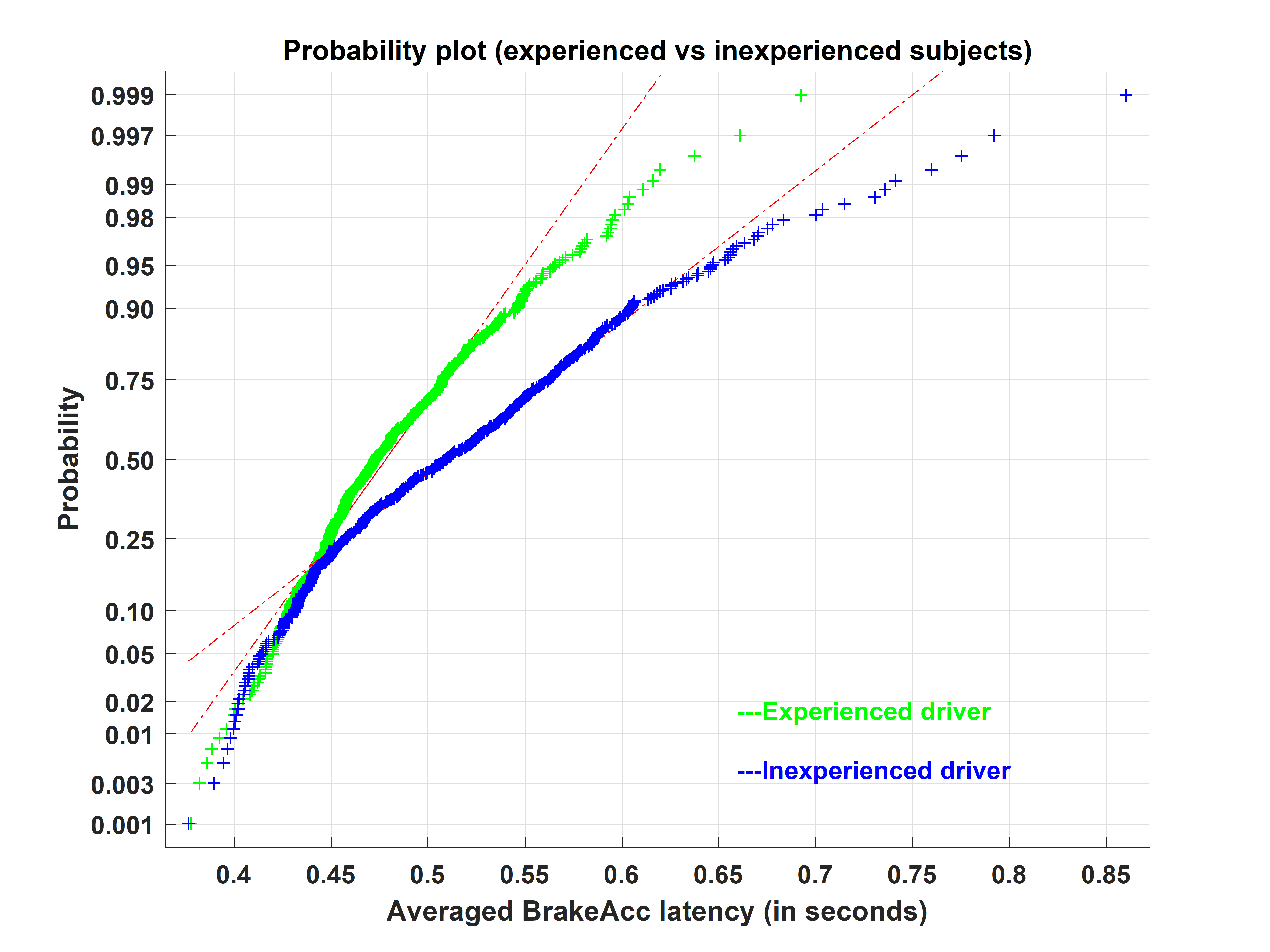}
\caption{Probability plot (experienced vs inexperienced subjects).}
\label{Fig16a}
\end{figure}
\begin{figure}[H]
\centering
\includegraphics[width=3.0in,height=3.0 in,keepaspectratio]{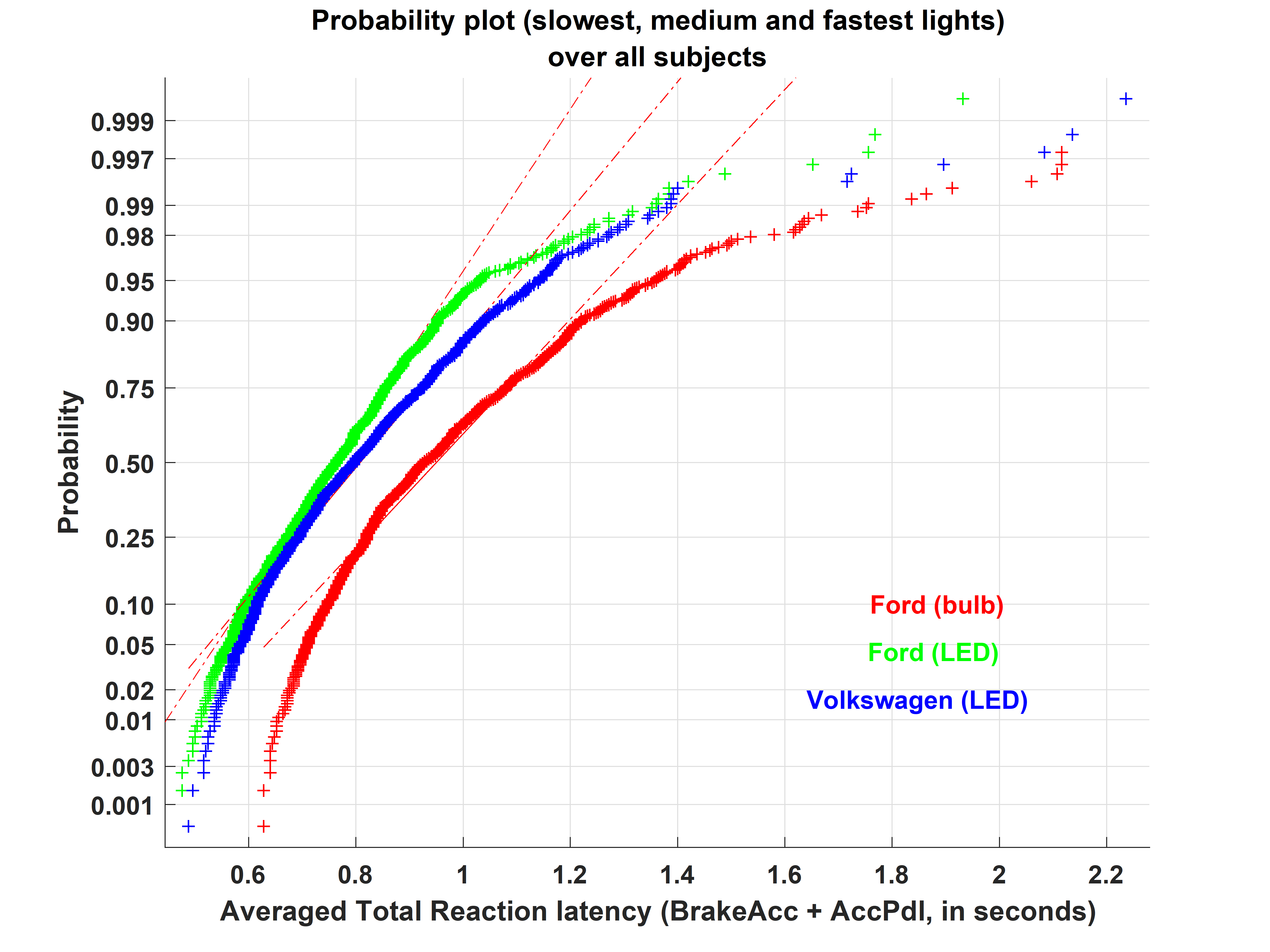}
\caption{Probability plot (slowest, medium and fastest lights) over all subjects.}
\label{Fig17a}
\end{figure}

The full bars of Figure \ref{Fig13a}) present the total reaction timings for all subjects. \textit{BrakePdl} for both bulbs was statistically slower than for any of the LED lights ( all pairwise cases $p<1e{-53}$). 
Among the LED units, the Volkswagen was slower than the Audi, Ford, Honda, Alfa Romeo and Nissan units (all pairwise cases $p<5e{-3}$) while the Mercedes unit was slower than the Ford ($U=-4.53, p<2.89e{-6}$). 
We speculate the results from the Volkswagen unit was at least partially a result of pattern in which it illuminates (see Section~\ref{sec:conclusion}).

The results also showed that \textit{BrakeAcc} is statistically longer than \textit{AccPdl} for every brake light (all pairwise $p=0$, indicating that it took longer for subjects $(0.50\pm0.05 s)$ to act on the detected brake light illumination than to depress the brake pedal $(0.33 \pm 0.10 s)$.
This indicated that more time was required by subjects to perceive the activation of brake lights, but they are generally quicker to act once brake light activation is recognised. 
Among the LED lights, the best and worst responses were mixed for each subject as shown in Table \ref{table:Table 4}. 
Nevertheless, the results do indicate that the time between seeing the brake light illuminating, and releasing the accelerator, is the critical interval where the different types of lights can influence the speed of braking reaction.

Figure \ref{Fig14a} shows the \textit{BrackAcc }latency versus the age of all the subjects (in years). 
There was no significant correlation ($r^2=0.0385, p=3.82e{-1}$), indicating clearly that age, within the range tested, had no influence on the speed of recognition of the brake light activation. 

Figure \ref{Fig15a} plots \textit{AccPdl }latency versus the experience of all the subjects (in months). 
There was no significant correlation statistically ($r^2=0.15, p=7.52e{-2}$), although the small $p$ value and $r^2=0.15$ do indicate that there is some correlation between driving experience and speed of recognition of the brake light activation, i.e. more experienced subjects are quicker to respond. 

There was no significant correlation statistically  when comparing \textit{AccPdl} latencies with age ($r^2=0.0164, p=5.70e{-1}$) showing that age does not have an influence on the reflex action. There was no significant correlation statistically when comparing \textit{AccPdl} latencies with experience ($r^2=0.0648, p=2.53e{-1}$) showing that age and experience do not have an influence on the reflex action, which could likely be more influenced by the subject's physical ability and innate speed of reflex movements. 

The probability distributions for experienced and inexperienced subjects using averaged \textit{BrakeAcc} latencies are shown in Figure \ref{Fig16a}.  The dotted red lines indicate the normal distribution and it can be seen that there is greater variation for inexperienced subjects (shown with a less steep red line). 
For example at $0.95$ probability ($5\%$), we can see that experienced subjects took an average $0.56$s to release the brake pedal while inexperienced subjects took an average of $0.65$s.    

Considering the three brake lights (slowest light overall, slowest LED and fastest light overall), Figure \ref{Fig17a} shows the probability distribution for all the subjects in terms of total reaction latencies (\textit{BrakePdl}).  
At $0.95$ probability ($5\%$), the latencies were $1.03$, $1.14$ and $1.36$s for the Ford Bulb, Volkswagen LED and Ford LED unit.  Considering the fastest speed of $1.03$s, the probability would stand at $0.89$ and $0.68$ for the Volkswagen LED and Ford Bulb respectively. Thus, $6\%$ more subjects were slower when comparing the Volkswagen and Ford LED lights and $27\%$ more subjects were slower when comparing the Ford bulb and Ford LED.

\begin{figure}[H]
\centering
\includegraphics[width=3.2in,height=3.2 in,keepaspectratio]{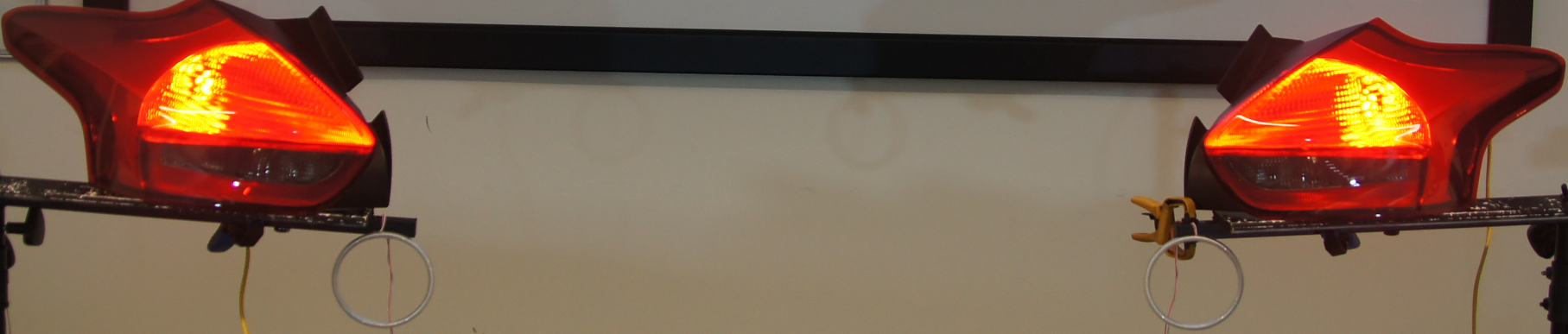}
\caption{The tested Ford brake light units.}
\label{Fig18a}
\end{figure}

\begin{figure}[H]
\centering
\includegraphics[width=3.2in,height=3.2 in,keepaspectratio]{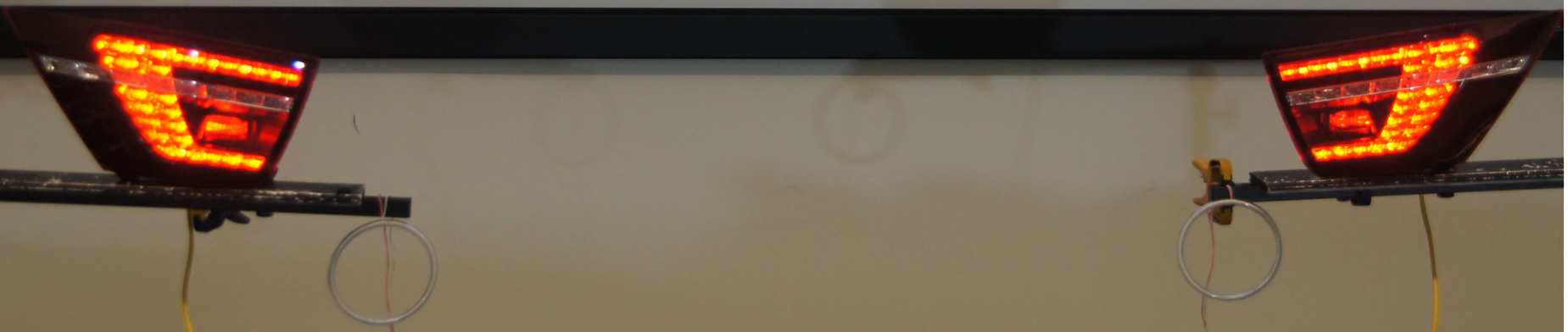}
\caption{The tested Volkswagen brake light units.}
\label{Fig19a}
\end{figure}

\section{Conclusion}
\label{sec:conclusion}

Reaction time data from $22$ subjects for ten brake light assemblies were analysed statistically. 
Results indicate that versions of the brake lights containing incandescent bulbs (e.g. Ford and Fiat) induced statistically slower reaction times than all of the tested LED units. 
It is known that incandescent bulbs take longer to illuminate (generally no discernible optical output for around $50$\,ms post switch on), but the cognitive reaction time delay difference was found to be about $170$\,ms between the incandescent bulb and LED equivalents (e.g. between the Ford LED and bulb assemblies).
This clearly reveals that LED units have the potential to evoke brain responses quicker.

 It was also shown that experienced subjects were quicker to realise the activation of a brake light, and hence release the accelerator quicker. A noteworthy finding here is that the brake light type also influenced the time between accelerator release and brake pedal depression. Furthermore, experienced subjects did not always act quicker than inexperienced subjects in this regard. These points are probably worthy of further analysis from the cognitive perspective, especially in terms of the relationship between shape and cognition.

The Ford brake light shell (Figure \ref{Fig18a}) had a larger lit area than the other brake lights, which could have led to improved visibility. The Volkswagen brake light (Figure \ref{Fig19a}) had a unique dispersed illumination pattern, with the major lit area being towards the exterior and less focused to the centre of the brake light unit. 
The Mercedes brake light (Figure \ref{Fig19a}) also had an elliptical illumination pattern, with the centre of the light unit being unlit. Comparing the lights inducing the slowest response (the bulb units), both lacked illumination at the centre of the brake shells, which could contribute to the slower times.

For our future work, we are planning to analyse the actual cognitive responses from the braking events using electroencephalogram (EEG) signals as this would allow us to understand the brain processes involved in the recognition of the lights and the corresponding braking actions. 
We will also be exploring running the experiments in real-life traffic conditions (i.e. live, on the road) to assess any deviation from the responses obtained in the laboratory environment. 

\section*{Acknowledgment}

We acknowledge the support of the Road Safety Trust (RST 90\_4\_18) in funding this study. The research was conducted with the aim of increasing road safety, and did not have the involvement of any manufacturer. We note that only one type of light assembly from a single vehicle within each manufacturers range was assessed, and results are therefore not to be interpreted as applying to any other vehicles or brake light assemblies, and certainly not as any endorsement or otherwise of a particular manufacturer. We also acknowledge that the sample sizes here are quite small, so it would be appropriate for replications of the work to confirm the findings.

\bibliographystyle{IEEEtran}

\bibliography{Brakelight}


\vspace*{-15mm}
\begin{IEEEbiography}[{\includegraphics[width=1in,height=1.25in,clip,keepaspectratio]{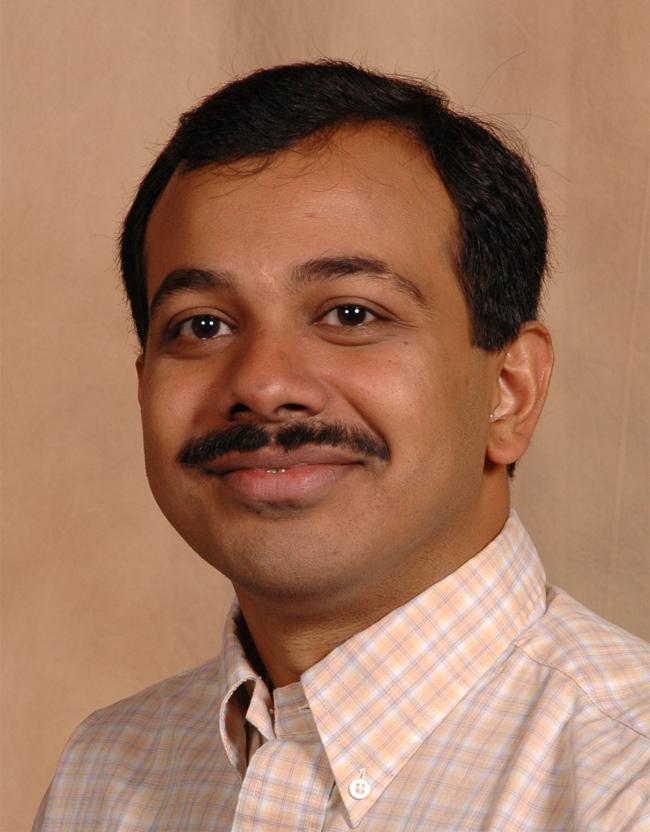}}] {Ramaswamy Palaniappan}

is currently a Reader in the School of Computing, University of Kent. His current research interests include signal processing and machine learning for electrophysiological applications. To date, he has written two text books in engineering and published over 200 papers (with over 3000 citations) in peer-reviewed journals, book chapters, and conference proceedings. He is a senior member of the Institute of Electrical and Electronics Engineers, member in Institution of Engineering and Technology, and Institute of Physics and Engineering in Medicine. He serves in editorial boards  for several international journals. He also serves in the prestigious Peer Review College for UK Research Councils and many other international grant funding bodies. He has supervised more than half a dozen postgraduate students to completion and has more than two decades of multi-disciplinary teaching experience in computer science and engineering (electrical and biomedical) disciplines. His pioneering work on revolutionary new areas of brain-computer interfaces and emerging biometrics has not only received international awards and recognition by the scientific community but also from the media and public. 
\end{IEEEbiography}

\vspace*{-20mm}
\begin{IEEEbiography}[{\includegraphics[width=1in,height=1.25in,clip,keepaspectratio]{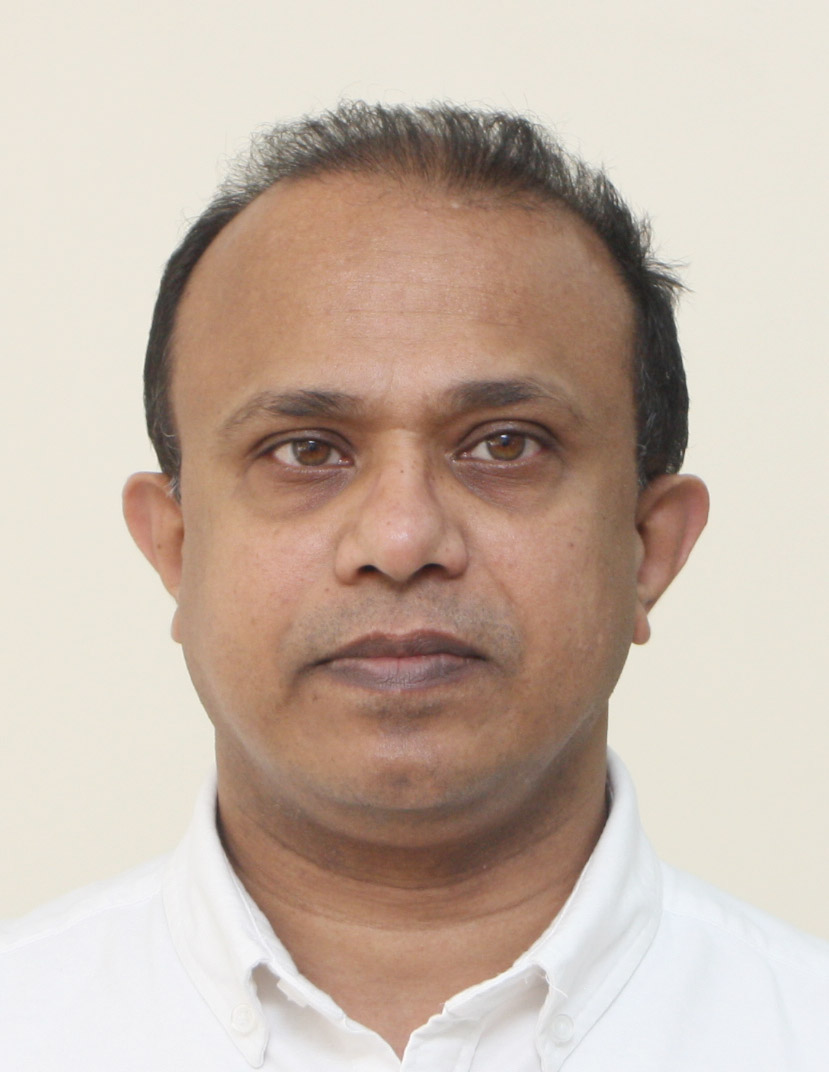}}] {Surej Mouli}

received his PhD degree in Computer Science from University of Kent, UK in 2017. He is currently working as a Research Associate at School of Computing, University of Kent. He is a senior member in IEEE, member of British Computer Society and IET. His main research interest includes IoT, Brain-computer Interfaces, EEG signal analysis, and Vagus Nerve Stimulation.

\end{IEEEbiography}

\vspace*{-10mm}
\begin{IEEEbiography}[{\includegraphics[width=1in,height=1.25in,clip,keepaspectratio]{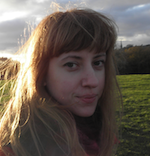}}] {Evangelina Fringi}

received her BSc in Mathematics from the National and Kapodistrian University of Athens, Greece in 2010 and an MA in Philosophy of Mind and Cognitive Science from the University of Birmingham, UK in 2012. She received her PhD degree in Computer Science from University of Birmingham, UK in 2020 working on Disney Research (computer speech recognition errors for children’s speech). She worked as a Research Associate at School of Computing, University of Kent and Research Assistant in Sensory Motor Neuroscience lab (SyMoN) in the University of Birmingham. Her main research interests include speech processing and social psychology.
\end{IEEEbiography}

\vspace*{-50mm}
\begin{IEEEbiography}[{\includegraphics[width=1in,height=1.25in,clip,keepaspectratio]{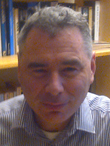}}] {Howard Bowman} Professor Bowman completed a PhD and Postdoc at Lancaster University and then took up a lectureship in Computer Science at the University of Kent at Canterbury. In this first phase of his career he developed mathematical theories of computation, particularly in concurrency theory. He was appointed to a chair at Kent in 2006. He now holds part-time positions at Kent and at Birmingham, with his work now exclusively focused on Cognitive Neuroscience. Professor Bowman applies the methods of Cognitive Neuroscience, especially EEG and Neural Modelling, to understanding a spectrum of Cognitive phenomena, including conscious perception, temporal attention and subliminal search. Much of his work focuses on verifying the simultaneous Type/ Serial Token theory of temporal attention and working memory encoding, which he developed with Brad Wyble.
\end{IEEEbiography}

\vspace*{-50mm}
\begin{IEEEbiography}
[{\includegraphics[width=1in,height=1.25in,clip,keepaspectratio]{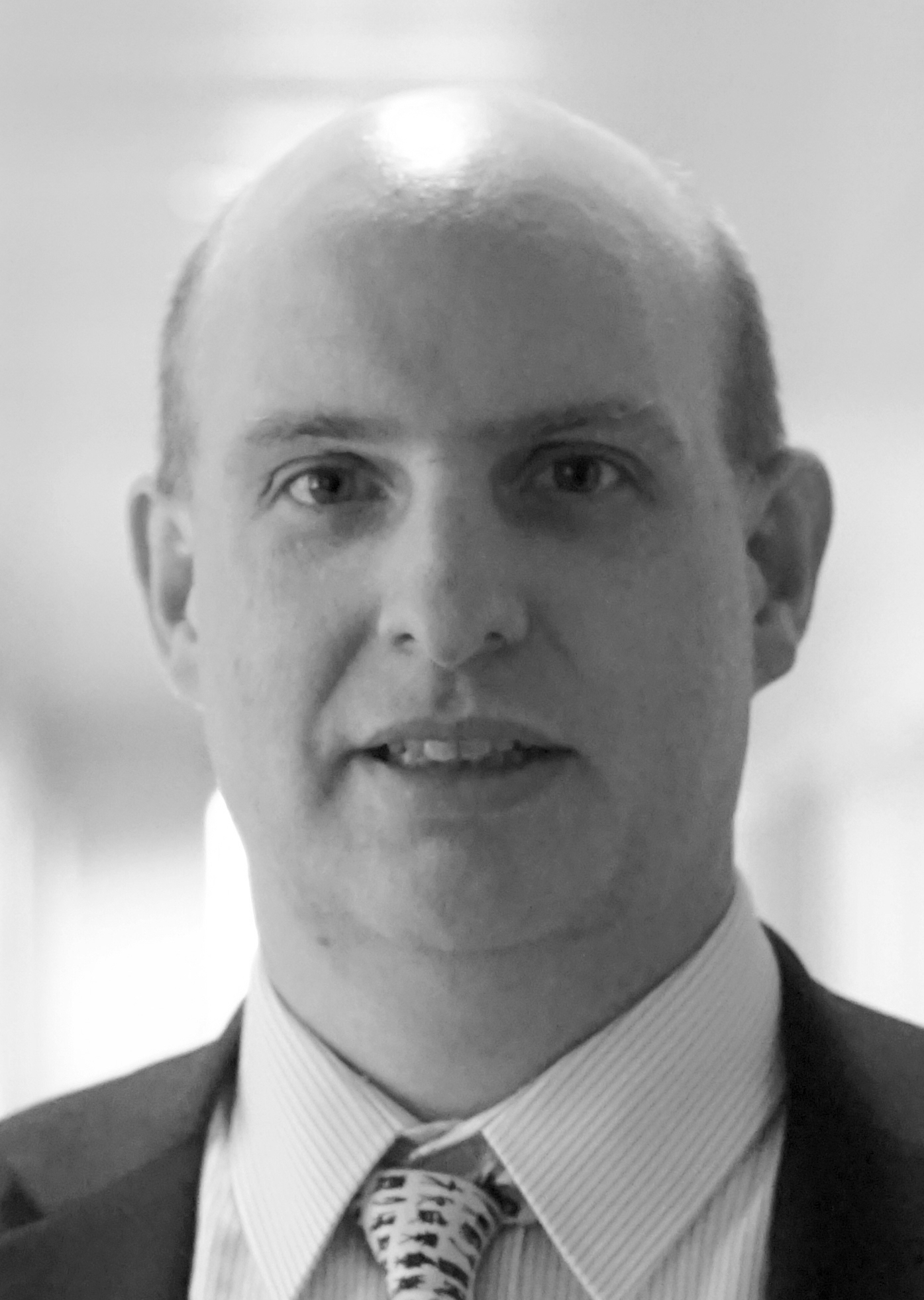}}]{Ian McLoughlin} (M'94, SM'04) completed his PhD in Electronic and Electrical Engineering at the University of Birmingham, UK in 1997. He has worked for over 10 years in the R\&D industry and almost 20 years in academia, on three continents. He is a professor at Singapore Institute of Technology, a visiting Professor at the University of Science and Technology of China, and is a Fellow of the IET. He has authored many papers, has several patents and has written four textbooks (so far) on speech processing and embedded computation.
\end{IEEEbiography}

\end{document}